\documentclass[aps,prl,twocolumn,superscriptaddress,raggedbottom,showpacs,floatfix]{revtex4-1}
\usepackage{color}
\usepackage{graphicx,latexsym}
\usepackage{epstopdf}
\usepackage{comment,bm}
\usepackage{amsmath}
\usepackage{amssymb}
\usepackage{mathtools}
\usepackage{graphics}
\usepackage{comment}
\usepackage{tikz}
\usepackage{mathtools}

\DeclarePairedDelimiter\floor{\lfloor}{\rfloor}

\newcommand{\beq}{\begin{equation}}
\newcommand{\eeq}{\end{equation}}
\newcommand{\bei}{\begin{itemize}}
\newcommand{\eei}{\end{itemize}}
\newcommand{\ben}{\begin{enumerate}}
\newcommand{\een}{\end{enumerate}}

\newcommand{\be}{{\mathbf e}}

\usepackage{hyperref}
\hypersetup{
   pdfpagemode=None, 
   pdfstartpage=1,
   pdfmenubar=true,
   pdftoolbar=true,
   colorlinks = true,
   linkcolor=blue,
   citecolor=blue,
   urlcolor=blue,
   bookmarksopen=false
 }

\definecolor{darkblue}{rgb}{0.,0.24,0.51}
\definecolor{britishracinggreen}{rgb}{0.0, 0.26, 0.15}
\definecolor{darkgreen}{rgb}{0,0.60,.2}

\def\be{\begin{equation}}
\def\ee{\end{equation}}

\begin{document}
\renewcommand{\vec}{\mathbf}
\renewcommand{\Re}{\mathop{\mathrm{Re}}\nolimits}
\renewcommand{\Im}{\mathop{\mathrm{Im}}\nolimits}

\title{Equatorial magnetoplasma waves}

\author{Cooper Finnigan}
\affiliation{School of Physics and Astronomy, Monash University, Victoria 3800, Australia}

\author{Mehdi Kargarian}
\affiliation{Department of Physics, Sharif University of Technology, Tehran 14588-89694, Iran}

\author{Dmitry K. Efimkin}
\email{dmitry.efimkin@monash.edu}
\affiliation{School of Physics and Astronomy, Monash University, Victoria 3800, Australia}
\affiliation{ARC Centre of Excellence in Future Low-Energy Electronics Technologies, Monash University, Victoria 3800, Australia}

\begin{abstract}
Due to its rotation, Earth traps a few equatorial ocean and atmospheric waves, including Kelvin, Yanai, Rossby, and Poincar\'e modes. It has been recently demonstrated that the mathematical origin of equatorial waves is intricately related to the nontrivial topology of hydrodynamic equations describing oceans or the atmosphere. In the present work, we consider plasma oscillations supported by a two-dimensional electron gas confined at the surface of a sphere or a cylinder. We argue that in the presence of a uniform magnetic field, these systems host a set of equatorial magnetoplasma waves that are counterparts to the equatorial waves trapped by Earth. For a spherical geometry, the equatorial modes are well developed only if their penetration length is smaller than the radius of the sphere. For a cylindrical geometry, the spectrum of equatorial modes is weakly dependent on the cylinder radius and overcomes finite-size effects. We argue that this exceptional robustness can be explained by destructive interference effects. We discuss possible experimental setups, including grains and rods composed of topological insulators (e.g., $\hbox{Bi}_2\hbox{Se}_3$) or metal-coated dielectrics (e.g., $\hbox{Au}_2 \hbox{S}$).
\end{abstract}

\date{\today}
\maketitle
\section{I. Introduction}
\label{SecI}
Over the last decade, topological states of matter (e.g., topological insulators, Weyl semimetals, and superconductors) have been a central topic in condensed matter physics~\cite{WeylReview,TopologicalInsulatorsReview,TopologicalSuperconductprsReview}. The unconventional and topologically protected surface states hosted by these materials have received much attention because of their excellent prospects for energy-efficient electronics, and spintronic devices.  More recently, the concept of topology has been fruitfully extended to other fields, including photonics~\cite{TopologicalPhotonicsReview1,TopologicalPhotonicsReview2,TopologicalPhotonicsReview3}, electrical circuits~\cite{TopologicalCircuits1,TopologicalCircuits2}, and non-Hermitian systems~\cite{NonHermitianTopology1,NonHermitianTopology2,NHadditional}.  

As another prominent success, the application of topology has enabled an explanation of the mathematical origin of equatorial ocean and atmospheric waves trapped at the Earth’s equator due its rotation~\cite{EquatorialClasssics,BookVallis06}. It has been recently revealed that the presence of the Coriolis force in the hydrodynamic equations not only opens a gap in the spectrum of ocean and atmospheric waves, but the latter can be classified as topologically nontrivial~\cite{TopologyEquatorialWaves}. The directions of the Coriolis force and therefore the topological number have opposite signs in the northern and southern hemispheres. This change in direction across the equator guarantees the presence of two chiral topologically protected trapped waves (Kelvin and Yanai) coexisting with trivial trapped modes (Rossby and Poincar\'e). As the mechanism for the formation of equatorial waves is quite generic, a question naturally arises regarding whether any equatorial waves can be engineered and probed in condensed matter or cold atom setups.

\begin{figure}[b]
	\begin{center}
	\vspace{-0.1in}
		\includegraphics[trim=0cm 0cm 0cm 0.2cm, clip, width=0.55\columnwidth]{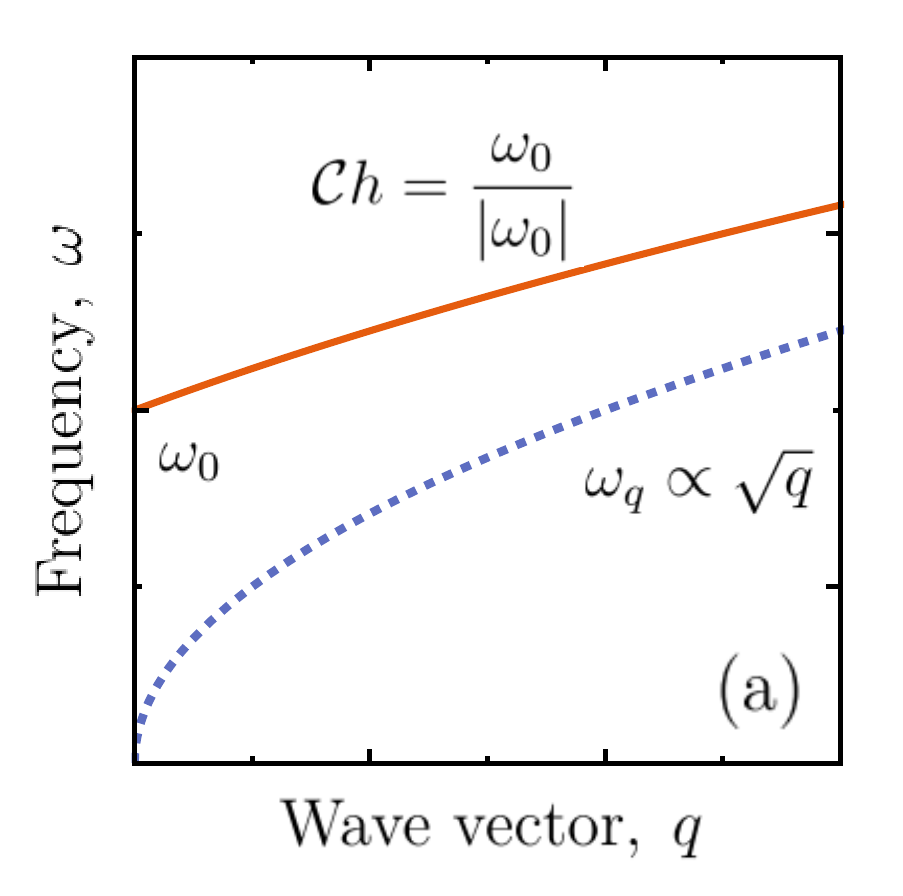} \quad
		\includegraphics[trim=1.5cm 8cm 22cm 4cm, clip, width=0.35\columnwidth]{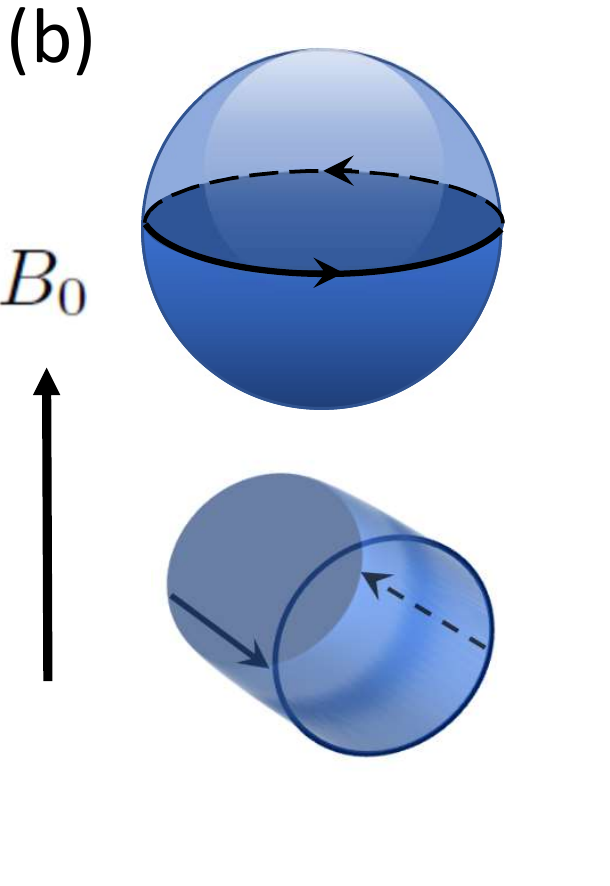}
		\caption{ (a) Dispersion of MP waves in the absence (blue dashed) or presence (solid red) of a magnetic field. In the latter case, the spectrum is not only gapped at the Larmor frequency $\omega_0$, but is also characterized by the topological Chern number $\mathcal{C}h=\omega_0/|\omega_0|$. (b) In the presence of a uniform magnetic field, a 2D electron gas confined at the surface of a sphere or a cylinder supports MP waves trapped along the equator or at opposite facets.}
		\label{Fig5}
	\end{center}
	\vspace{-0.2in}
\end{figure}

Hydrodynamic equations govern not only ocean and atmospheric waves, but also plasma waves. The latter are supported by an electron gas and represent coupled oscillations in the electron density and electric field.  Moreover, equations describing plasma waves in a two-dimensional (2D) electron gas on the top of the extended gate (e.g., in field-effect transistors) can be mapped~\cite{DiakonovShur,Chaplik1972PossibleCO} to shallow water hydrodynamics, which are usually employed to describe equatorial waves trapped by Earth~\cite{EquatorialClasssics,TopologyEquatorialWaves}. The role of the Coriolis force is played by the Lorentz force due to the external magnetic field. The magnetic field opens a gap in the spectrum of magnetoplasma (MP) waves~\footnote{Plasma waves in the presence of a magnetic field are usually referred to as magnetoplasma waves or magnetoplasmons}, as shown in Fig.~1-a, which results in a nontrivial topology ~\cite{MagnetoPlasmons1}.

Motivated by these observations, we consider plasma oscillations supported by a 2D electron gas confined at the surface of a sphere or cylinder. We argue that in the presence of a uniform magnetic field, these systems host equatorial MP waves. These waves are illustrated in Fig.~1-b and are the counterpart of ocean waves trapped by Earth. We investigate the dependence of their spectrum on the sphere and cylinder radius. For a spherical geometry, the spectrum of equatorial MP waves is sensitive to finite-size effects and is well developed only if the sphere radius exceeds the penetration length for the equatorial MP waves. For a cylindrical geometry, the spectrum of equatorial modes is weakly dependent on the cylinder radius and the overcomes finite-size effects. We argue that the exceptional robustness of this spectrum can be explained by destructive interference between interfacet couplings across the top and bottom hemicylinders. We discuss possible experimental setups, including grains and rods composed of topological insulators (e.g., $\hbox{Bi}_2\hbox{Se}_3$) or metal-coated dielectrics (e.g., $\hbox{Au}_2 \hbox{S}$).

The remainder of this paper is organized as follows. In Sec.~II, we discuss the primary equations describing MP waves supported by a 2D electron gas in a magnetic field. In Sec.~III, we reconsider edge MP modes localized at the domain wall, where the magnetic field switches its sign. Sec.~IV is devoted to equatorial MP modes in spherical and cylindrical geometries. Sec.~V presents discussions and conclusions.

\section{II. Magnetoplasma waves}
The long-wavelength behavior of MP waves supported by a 2D electron gas can be described with the help of classical hydrodynamic equations~\cite{Fetter1}. After linearization, equations for electron density $\rho_{t\vec{r}}$ and electric current $\vec{j}_{t\vec{r}}$ can be presented as    
\begin{align}
\label{Continuity}
&\partial_t \rho(\vec{r},t)+\vec{\nabla}\cdot \vec{j}( \vec{r},t)=0, \\ \label{Newton} &\partial_t \vec{j}(\vec{r},t)=\frac{ne^2}{m} \vec{E}(\vec{r},t) + \frac{e}{mc}[\vec{j}(\vec{r},t)\times \vec{B}(\vec{r})].
\end{align}
Here, $m$ is the cyclotron electron mass~\footnote{For electrons with conventional quadratic dispersion, this term is equal to their band mass. For electrons with unconventional but isotropic dispersion (e.g., with relativistic-like Dirac dispersion), the mass $m=\epsilon_\mathrm{F}/v_\mathrm{F}$ is given by the ratio between the Fermi energy $\epsilon_\mathrm{F}$ and Fermi velocity $v_\mathrm{F}$.}. The electric field     
$E(\vec{r},t)=-\nabla \phi(\vec{r},t)$ is not external, but is created by electron density oscillations and is treated in a self-consistent manner. The corresponding scalar potential $\phi(\vec{r},t)$ is given by 
\begin{equation}
\label{Potential}
\phi(\vec{r},t)=\int d\vec{r}' V(\vec{r}-\vec{r}') \rho(\vec{r}',t),\quad \; V(\vec{q})=\frac{2\pi}{q\kappa(\vec{q})},
\end{equation}
where $V(\vec{r}-\vec{r}')$ represents interparticle interactions and $V(\vec{q})$ gives the corresponding Fourier transform. Here, $\kappa(\vec{q})$ is the dielectric function of the medium surrounding the 2D electron gas. Its explicit wave vector dependence for experimentally relevant setups is specified  below. The system of equations, Eqs.~(\ref{Continuity}-\ref{Potential}), is valid for 2D electrons confined at the surface of an arbitrary geometry. However, it is instructive to start with a discussion of the spectrum and its topology for MP waves in a planar geometry. 
  
A key component to the topological classification of MP waves is the transformation reported in Ref.~\cite{MagnetoPlasmons1}, which recasts the system of equations, Eqs.~(\ref{Continuity}) - (\ref{Potential}), into a Hermitian-Schr\"{o}dinger-like eigenvalue problem, $\omega \psi(\vec{q},\omega)=\hat{H}(\vec{q}) \psi(\vec{q},\omega)$. Here, we have performed the Fourier transform and have introduced $\psi(\vec{q},\omega)=\{j^+(\vec{q},\omega),j^0(\vec{q}, \omega),j^-(\vec{q},\omega)\}$ with $j^\pm(\vec{q},\omega)=(j^x(\vec{q},\omega)\pm i j^y(\vec{q},\omega))/\sqrt{2}$ and $j^0(\vec{q},\omega)=\omega_\mathrm{p}(\vec{q}) \rho(\vec{q},\omega)/q$. The resulting  effective Hamiltonian $H(\vec{q})$ for the MP problem is given by 
\begin{equation}
\label{HamiltonianFull}
\hat{H}(\vec{q})=\begin{pmatrix}
\omega_0 & \frac{\omega_\mathrm{p}(\vec{q}) e^{i\phi_\vec{q}}}{\sqrt{2}}& 0 \\
\frac{\omega_\mathrm{p}(\vec{q}) e^{-i\phi_\vec{q}}}{\sqrt{2}} & 0 & \frac{\omega_\mathrm{p}(\vec{q}) e^{i\phi_\vec{q}}}{\sqrt{2}} \\
0 & \frac{\omega_\mathrm{p}(\vec{q}) e^{-i\phi_\vec{q}}}{\sqrt{2}} & - \omega_0 \\
\end{pmatrix}.
\end{equation}
Here, $\phi_\vec{q}$ is the polar angle for a wave vector $\vec{q}$, $\omega_0=eB_0/mc$ is the Larmor frequency for electrons in a uniform magnetic field $B_0$, and $\omega_\mathrm{p}(\vec{q})=\sqrt{2\pi n e^2 q /m \kappa(\vec{q})}$ is the dispersion for plasma waves in the absence of a magnetic field. The classical nature of the underlying problem manifests in the presence of a particle-hole symmetry, $C H(\vec{q})C^{-1}=-H(-\vec{q})$~\footnote{The explicit expression for $C$ is given by 
$$C=
\begin{pmatrix}
0 & 0 & \mathcal{K} \\
0 & \mathcal{K} & 0 \\
\mathcal{K} & 0 & 0 \\
\end{pmatrix},
$$
where $\mathcal{K}$ is the complex conjugation operator.}. The states connected by the transformation $C$ have opposite frequencies and are not independent. In addition, the symmetry guarantees that any observables (e.g., $\rho(\vec{r},t)$ or $j(\vec{r},t)$ are real numbers. The eigenvalues of the Hamilton $H(\vec{q})$ are given by
\begin{equation}
\label{MPBulkDispersion}
\Omega^\pm(\vec{q})=\pm \Omega(\vec{q}), \quad \hbox{and} \quad \Omega^0(\vec{q})=0. 
\end{equation}
The positive-frequency branch $\Omega(\vec{q})= \sqrt{\omega_0^2+\omega_\mathrm{p}^2(\vec{q})}$ governs the dispersion relation for MP waves, as presented in Fig~1-a. In the presence of a magnetic field, the dispersion acquires the gap given by the Larmor frequency $\omega_0$. The negative-frequency branch is connected to the positive branch by the particle-hole symmetry transformation and is not dynamically independent. The interplay of the inversion ($\Omega^0(-\vec{q})=\Omega^0(\vec{q})$) and particle-hole ($\Omega^0(-\vec{q})=-\Omega^0(\vec{q})$) symmetries dictates that $\Omega^0(\vec{q})=0$, which causes the zero-energy branch to be spurious.

By recasting the equations, Eqs.~(\ref{Continuity}- \ref{Potential}), as a Hermitian eigenvalue problem, the topology of the MP spectrum can be classified~\cite{MagnetoPlasmons1}. Because $H(\vec{q})$ belongs to the D-class~\cite{TopClassification}, each branch can be characterized by the Chern number as $\mathcal{C}h_{\pm}=\mathrm{sgn}[\omega_0]$ with $\mathcal{C}h_{0}=0$. The nontrivial topology can be tracked by presenting the effective Hamiltonian $H(\vec{q})$ as $H(\vec{q})=\vec{h}(\vec{q}) \cdot \bm{\sigma }$, where $\bm{\sigma}$ is a spin-1 generalization of the Pauli matrix set~\footnote{Explicit expressions for the components of spin-1 Pauli matrices $\bm{\sigma}$ are given by 

\begin{equation*}\begin{split}
\sigma_x=
\begin{pmatrix}
0 & \frac{1}{\sqrt{2}} & 0 \\
\frac{1}{\sqrt{2}} & 0 & \frac{1}{\sqrt{2}} \\
0 & \frac{1}{\sqrt{2}} & 0 \\
\end{pmatrix}, \quad \quad \sigma_y=
\begin{pmatrix}
0 & - \frac{i}{\sqrt{2}} & 0 \\
\frac{i}{\sqrt{2}} & 0 & - \frac{i}{\sqrt{2}} \\
0 & \frac{i}{\sqrt{2}} & 0 \\
\end{pmatrix}, \\ 
 \sigma_z=
\begin{pmatrix}
1 & 0 & 0 \\
0 & 0 & 0 \\
0 & 0 & -1 \\
\end{pmatrix}.\quad \quad \quad \quad \quad \quad \quad \quad 
\end{split}
\end{equation*}}
 and $h(\vec{q})=\{\omega_\mathrm{p}(\vec{q}) \cos\phi_{\vec{q}}, -\omega_\mathrm{p}(\vec{q}) \sin\phi_{\vec{q}}, \omega_0\}$. The unit vector $n(\vec{q})=\vec{h}(\vec{q})/h(\vec{q})$ follows the meron-like texture and spans half of the Bloch sphere. In practice, the vector points up ($\omega_0>0$) or down ($\omega_0<0$) at $\vec{q}\rightarrow0$ and has a vortex-like in-plane texture at $\vec{q}\rightarrow\infty$. The topology of the MP wave spectrum is insensitive to the details of screening by external media encoded in $\kappa(\vec{q})$, but the latter shapes the dispersion of MP waves.    

For a 2D electron gas embedded in dielectric media (e.g., at the interface between air and an insulating substrate), the dielectric function can be approximated as the wave-vector-independent $\kappa$. The resulting dispersion of plasma waves in the absence of an external magnetic field shows a square-root dependence, $\omega_\mathrm{p}(\vec{q}) \propto \sqrt{q}$, which reflects the nonlocal nature of long-range Coulomb interactions~\footnote{The motion of the electrons is confined to the plane, but interactions are mediated by an electric field that is extended in three-dimensional space.}. Once it is transformed to real space, the square-root dispersion does not have any local representation in terms of $\nabla_\vec{r}$, which complicates analytical treatments for MP edge states. For a 2D electron gas placed at a distance $L$ from an extended gate, the charge carriers in the gate actively participate in screening interactions. The corresponding wave vector dependence of the dielectric constant can be approximated as $\kappa(\vec{q})=\kappa /\tanh{q L}$~\cite{Fetter1}. The long-range nature of the interactions is lost, and the plasma wave dispersion    
$\omega_\mathrm{p}(\vec{q})=\sqrt{2\pi n e^2 q\; \mathrm{th}(q L) /m \kappa}$ becomes linear $\omega_\mathrm{p}(\vec{q})\approx u q$ at long wavelengths,  $q L\lesssim1$. Here, $u=\sqrt{2\pi n e^2 L/m \kappa}$ is the corresponding plasma wave velocity. The effective Hamiltonian $H_\vec{q}$ is simplified as 
\begin{equation}
\label{HamiltonianFullPrime}
\hat{H}'(\vec{q})=\begin{pmatrix}
\omega_0 & \frac{u (q_x+iq_y)}{\sqrt{2}}& 0 \\
\frac{u (q_x-iq_y)}{\sqrt{2}} & 0 & \frac{u (q_x+iq_y)}{\sqrt{2}} \\
0 & \frac{u (q_x-iq_y)}{\sqrt{2}} & - \omega_0 \\
\end{pmatrix}
\end{equation}
and can be transformed in real space in a straightforward manner. It should be mentioned that in this approximation, there is a one-to-one mapping between plasma waves and surface waves within the shallow-water hydrodynamics usually used for describing equatorial ocean waves trapped by Earth~\cite{TopologyEquatorialWaves}.   

The nontrivial topology of the bulk MP wave spectrum dictates the presence of edge MP modes localized at sample boundaries or domain walls, where the magnetic field flips its sign. Before addressing the equatorial MP waves in spherical and cylindrical geometries, which are the focus of Sec.~IV, we reconsider the domain wall problem in planar geometry and demonstrate that the spectrum of edge MP modes is more complicated than was previously reported~\cite{MagnetoPlasmons1}.   

\section{III. Domain walls in planar geometry}
According to the bulk-edge correspondence, the magnetic field domain wall is expected to host a pair of chiral states propagating in only one direction. The domain wall problem has already been considered in Ref.~[\onlinecite{MagnetoPlasmons1}], in which the plasma wave spectrum is assumed to be linear (interactions are overscreened) and a step-like ansatz $b(x)=B_0\hbox{sgn}[x]$ is employed for the magnetic field profile.  With these approximations, the spectrum includes bulk bands given by Eq.~(\ref{MPBulkDispersion}) as edge MP modes. The positive-frequency region of the spectrum for edge MP waves is given by 
\begin{equation}
\label{SpectrumDW1}
\Omega^{\mathrm{I}}(k)=-u k, \quad \quad \Omega^{\mathrm{II}}(k)=\omega_0  \quad \hbox{at} \quad k>0.
\end{equation}
Here, $k$ is the wave vector along the domain wall. The first mode $\Omega^{\mathrm{I}}(k)$ is chiral and connects the positive and zero-frequency branches, but this is not the case for the second mode $\Omega^{\mathrm{II}}(k)$. This unusual behavior can be regularized by introducing transverse viscosity into the hydrodynamic equations~\cite{ABEC1}. However, it has been further discovered that the spectrum is sensitive to boundary conditions at the domain wall and to the regularization scheme, which can be referred to as an \emph{anomalous} bulk- or interface-boundary correspondence~\cite{ABEC1,ABEC2,ABEC3,ABEC4,ABEC5,BulkEdgeAdditional}. 

%This anomalous behavior is also expected to be an artifact of the employed hard-wall domain wall ansatz, which has been used because it admits a straightforward analytical solution. 

In the present paper, we reconsider the domain wall problem. We demonstrate that the edge MP wave spectrum admits an analytical solution for the smooth ansatz $B(x)=B_0 b(x)$ with $b(x)=\tanh(x/d)$. Here, $d$ is the domain wall width. We start with the exclusion of $j^0(\vec{r},t)$ and incorporate translational symmetry along the domain wall. The latter allows us to search for solutions in the following form: 
\begin{equation}
\label{WaveFunctionAnzatz}
\begin{split}
\vec{j}^{x(y)}(\vec{r},t)=J^{x(y)}(x) e^{i (k y-\omega t)}, \\ j^{0}(\vec{r},t)=J^{0}(x) e^{i (k y-\omega t)}.
\end{split}
\end{equation}
As a result, $J^{x(y)}\equiv J^{x(y)}(x)$ satisfy the following set of equations: 
\begin{equation}
\label{EigenvalueProblem1}
\begin{split}
(\omega^2+u^2\partial_x^2)J^x=-i (u^2 k\partial_x-\omega_0 \omega b(x)) J^y, \\
(\omega^2-u^2 k^2)J^y=-i (u^2 k\partial_x+\omega_0 \omega b(x)) J^x.
\end{split}
\end{equation}
As clearly shown, these equations admit a pair of solutions:
\begin{equation}
\label{KelvinMode}
 \quad  J^x=0, \quad \quad J^y\sim \exp\left[\pm\frac{\omega_0}{u} \int_0^x dx' b(x')\right]
\end{equation}
with the dispersion $\Omega_\mathrm{K}^{\pm}(k)=\pm u k$.  
For the considered ansatz $b(x)=\tanh(x/d)$, only mode  $\Omega_\mathrm{K}^{-}(k)=- u k$ can be properly normalized and matches with $\Omega^{\mathrm{I}}(k)$ given by Eq.~(\ref{SpectrumDW1}). If we follow the terminology for equatorial waves, the mode 
$\Omega_\mathrm{K}(k)\equiv \Omega_\mathrm{K}^{-}(k)$ can be referred to as the Kelvin MP mode. Importantly, its dispersion does not depend on the details of the domain wall profile; rather, it is only required that the latter has a kink~\footnote{The kink profile implies that $b(-\infty)<0$ and $b(\infty)>0$. It should also be noted that the mathematical structure of this solution is reminiscent of the celebrated Jackiw-Rebbi solution for the massive Dirac model with position-dependent mass~\cite{JackiwRebbi}.}. As a longitudinal edge wave ($J^x=0$), the Kelvin mode is accompanied by density oscillations with the same profile $J^0=-J^y$ as for $J^y$. For the employed ansatz $b(x)$, the profile is symmetric, 
$J^y\propto1/[\cosh\left(x/d\right)]^{\omega_0 d/v}$, across the domain wall.

If $\omega^2\neq u^2 q^2$, the set of equations in Eq.~(\ref{EigenvalueProblem1}) can be combined in a single equation given by 
\begin{equation*}
\label{Eigenvalue2}
-u^2 \partial_x^2 J^x+\left[u^2k^2+\omega_0^2 b^2(x)-\omega^2- \frac{u^2k \omega_0}{\omega} \partial_x b(x)\right] J^x=0.    
\end{equation*}
If we incorporate the explicit expression for the domain wall profile $b(x)=\tanh(x/d)$ and re-scale the coordinate $x\rightarrow x/d$, this  eigenvalue problem transforms into a one-dimensional quantum mechanical problem with the P$\ddot{o}$schl–Teller (PT) confining potential, which is given by   
\begin{equation}
-\partial_x^2 J^x-\frac{\lambda (\lambda+1)}{\cosh^2x}J^x=\epsilon J^x.    
\end{equation}
The parameter $\lambda$ and eigenenergy $\epsilon$ depend on both the wave vector $k$ and frequency $\omega$ and are given by   
\begin{equation}
\label{Eigenvalue3}
\begin{split}
\lambda (\lambda+1)=\frac{\omega_0}{\omega_\mathrm{d}} \frac{u k}{\omega}+ \frac{\omega^2_0}{\omega_\mathrm{d}^2}, \quad   \; 
\epsilon=\frac{\omega^2-u^2k^2-\omega_0^2}{\omega_\mathrm{d}^2}.
\end{split}
\end{equation}
Here, $\omega_\mathrm{d}=u/d$ is the frequency scale determined by the domain wall width. The PT potential belongs to the class of supersymmetric potentials~\cite{Susy1,Susy2}. Its spectrum has been extensively studied and  is known to admit an analytical solution.  The spectrum includes a continuum of extended states and a set of discrete states given by 
\begin{align}
\hbox{continuous:}& \quad &\epsilon>0,&\\
\label{Eigenvalue4}\hbox{discrete:}&\quad &\epsilon_n=-(\lambda-n)^2,& \quad n\le \lambda.    
\end{align}
The continuous region $\epsilon>0$ determines the continuum of bulk MP states with frequencies $\omega^2=\omega_0^2+v^2 k^2 +\epsilon\omega_\mathrm{d}^2$. The discrete bound states of the PT problem are intricately related to the MP edge modes trapped at the domain wall. The dispersion relation for edge MP waves satisfies the following equation, which can be obtained if we combine Eqs.~(\ref{Eigenvalue3}) and (\ref{Eigenvalue4}): 
\begin{equation}
\label{Eigenvalue5}
\begin{split}
\omega^2-(uk)^2+ \omega_0 \omega_\mathrm{d}\Big[\frac{uk}{\omega}- \\ \frac{\sqrt{\omega_0^2+u^2k^2-\omega^2}}{\omega_0}(2n+1)- \frac{\omega_\mathrm{d}}{\omega_0} n(n+1)\Big]=0.
\end{split}
\end{equation}
Due to its nonlinear nature, the equation can have multiple solutions for given $n=0,1,...$ and $k$. In dimensionless units $\omega/\omega_0$ and $k/k_0$ with $k_0=\omega_0/u$, the edge MP wave spectrum depends only on the parameter $\alpha=\omega_\mathrm{d}/\omega_0$. The spectrum for smooth ($\alpha=0.2$) and sharp ($\alpha=5$) domain wall profiles is presented in Fig.~2. If we follow the terminology for equatorial waves, the resulting MP edge modes can be labeled as Yanai, Poincar\'e, and Rossby modes. It is instructive to briefly discuss these modes separately.

\begin{figure}[t]
	\begin{center}
		\includegraphics[trim=1.2cm 0.3cm 0.9cm 1cm, clip, width=\columnwidth]{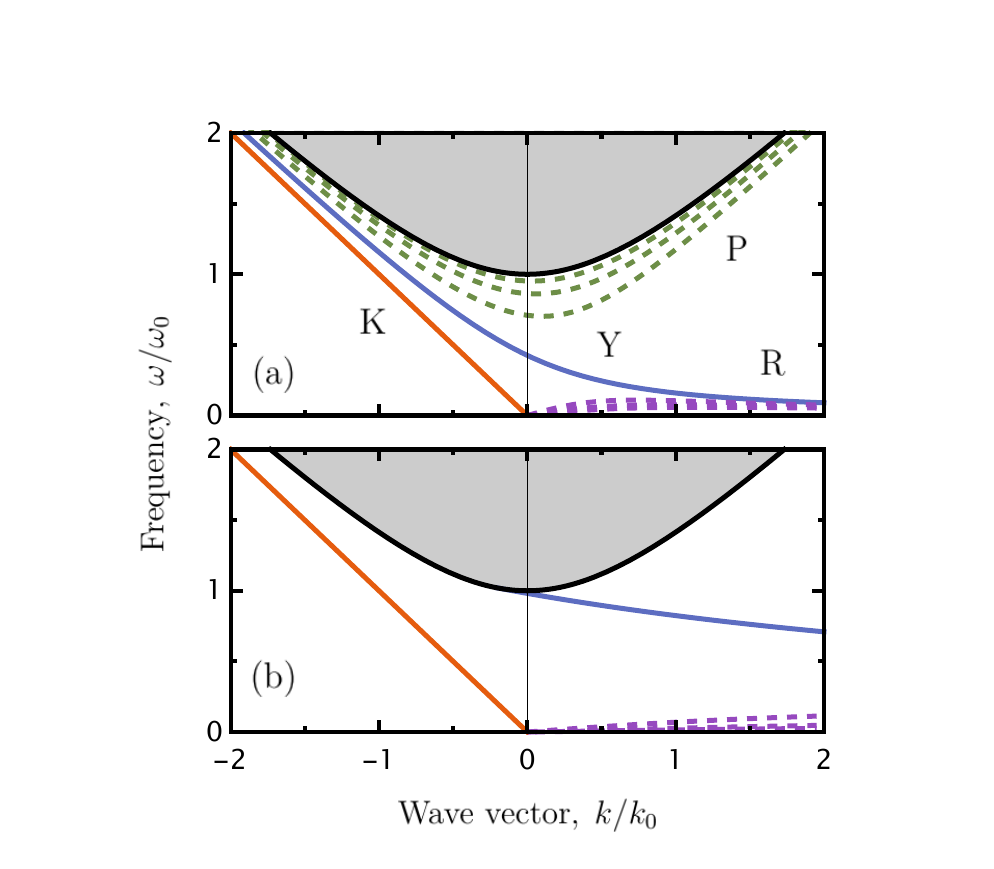}
		\caption{Spectrum for edge MP waves trapped by magnetic field domain walls with smooth ((a) $\alpha=0.2$) and sharp ((b) $\alpha=5$) profiles. The spectrum includes chiral and protected Kelvin (K) and Yanai (Y) modes coexisting with topologically trivial Rossby (R) and Poincar\'e (P) modes. The shaded area denotes the continuum of bulk extended states. }
		\label{Fig5}
	\end{center}
	\vspace{-0.2in}
\end{figure}

\emph{Yanai mode}. This mode is chiral and connects the bulk state continuum with zero-frequency states. The stability of the Yanai mode is intricately related to the fact that it originates from the ground bound state ($n=0$) for the PT problem~\footnote{At $k=0$, the prefactor $\lambda(\lambda+1)=\omega_0^2/\omega_\mathrm{d}^2$ is positive, which ensures the presence of at least one bound state trapped by the PT quantum well potential.}. Therefore, the profile of transverse current oscillations within the Yanai mode is symmetric and is given by $J^x\propto1/[\cosh\left(x/d\right)]^\lambda$. However, the profiles for the longitudinal current and density oscillations are antisymmetric across the domain wall, $J^y, J^0\propto b(x) J^x$, which ensures that wave functions for Kelvin and Yanai waves are orthogonal to each other. Their cumbersome expressions are presented in Appendix A.  %The frequency of the Yanai wave at $k=0$ is equal to  $\Omega^\mathrm{Y}_{k=0}=(-\omega_\mathrm{d}^2+\sqrt{\omega_\mathrm{d}^4+4 \omega_\mathrm{d}^2 \omega_0^2})/2)^{1/2}$. The frequency interpolates between $0$ and $\omega_0$ with increasing $\alpha$.         

\emph{Poincar\'e modes}. The Poincar\'e modes are trivial edge modes propagating in both directions along the domain wall. Poincar\'e modes originate from the excited bound states for the PT problem and are therefore labeled by the discrete index $n=1,2,3,...$. These states appear only if the domain wall is sufficiently smooth. Their number can be counted as $N_\mathrm{P}=\floor{(\sqrt{1+4 \omega_0^2/\omega_{\mathrm{d}}^2}-1)/2}$, and the first Poincar\'e mode splits from the bulk continuum at $\alpha=1/\sqrt{2}$~\footnote{We use $\floor{...}$ for the integer part of a real number.}. 

\emph{Rossby modes}. These low-energy modes result from the reshaping of zero-frequency bulk states~\footnote{In the presence of the domain wall, the inversion symmetry is broken, and these modes are no longer pinned at zero frequency.}. The Rossby modes also originate from the excited bound states for the PT problem and are therefore labeled by the discrete index $n=1,2,3,...$, similar to the Poincar\'e modes.
The dispersion of the Rossby modes is well approximated by 
\begin{equation*}
\Omega_{\mathrm{R}}^{n}(k)\approx\frac{u k \omega_\mathrm{d}}{u^2k^2+\omega_\mathrm{d}\sqrt{\omega_0^2+u^2k^2} (2n+1)+\omega_\mathrm{d}^2 n(n+1)}
\end{equation*}
and reaches a maximum at intermediate wave vectors for which the group velocity $\partial_k\Omega_{\mathrm{R}}^{n}(k)$ flips its sign. For this reason, these modes also propagate in both directions along the domain wall. The slowly varying spatial profile of the magnetic field is an important ingredient for the formation of Rossby MP modes. To the best of our knowledge, these modes have been previously overlooked in condensed matter setups. However, they have been well documented in stellar magnetohydrodynamics as well as in ocean and atmosphere hydrodynamics (see Ref.~\cite{RossbyReview} and references therein for a review). 

The presence of chiral Kelvin and Yanai modes is in agreement with the bulk-boundary correspondence, which dictates the presence of two topologically protected modes. Our calculations have clearly demonstrated that the unusual behavior of the spectrum, Eq.~(\ref{SpectrumDW1}), and the \emph{anomalous} bulk-edge correspondence are artifacts of the step-like ansatz for the magnetic domain wall profile.

It is instructive to consider the limit toward the domain wall with a step-like profile, $d\rightarrow 0$ (or $\alpha\rightarrow \infty$). In this limit, there are no Poincar\'e modes. The dispersion for the Kelvin mode is $\alpha$-independent, but the behavior of the Yanai and Rossby modes is quite sensitive.  For small wave vectors, $k\lesssim \alpha^{-1}$, the dispersion curves become flattened, $\Omega_{\mathrm{R}}^{n}(k)\approx0$ and $\Omega_\mathrm{Y}(k)\approx\omega_0$, mimicking the spectrum, Eq.~(\ref{SpectrumDW1}), derived for the step-like profile. However, for large wave vectors, $k\gtrsim  \alpha^{-1}$, both modes become dispersive and deviate from Eq.~(\ref{SpectrumDW1}). The lack of a smooth transition in the limit $d\rightarrow 0$ can be seen as another signature of the \emph{anomalous} bulk-boundary correspondence. 

The derivation presented above for the spectrum of edge MP modes assumes that that Coulomb interactions are overscreened. However, the classification of edge modes is generic and does not rely on the details of screening provided by external media. We will use this classification in the next section, in which spherical and cylindrical geometries are considered.  

\section{IV. Spherical and cylindrical geometries} \label{ConfinedGeometries}
The system of equations describing MP waves is valid for a 2D electron gas confined at the surface of an arbitrary geometry. Only the magnetic field component perpendicular to the surface influences the dynamics of electrons via the Lorentz force, which naturally becomes position-dependent for a curved surface. For a sphere or cylinder penetrated by a uniform magnetic field $B_0$, the radial component has a profile of $B_\mathrm{R}(\vec{r})=B_0 \cos (\theta)$, where $\theta$ is the corresponding polar angle. The radial component $B_\mathrm{R}
(\vec{r})$ vanishes and changes sign along the equator or at two facets, where the presence of equatorial MP modes is anticipated.

This setup can be naturally realized in grains and rods composed of a topological insulator (e.g., $\hbox{Bi}_2 \hbox{Se}_3$) that exhibit insulating bulk behavior~\footnote{We ignore the disorder-induced residual metallic bulk conductivity that usually appears in topological insulators, including $\hbox{Bi}_2 \hbox{Se}_3$ samples. For thin films with thickness $l\lesssim 0.3 \; \mu \hbox{m}$, the response has been argued to be fully dominated by the Dirac states.} and have topologically protected surface states. A fabrication of micron- and submicron-sized grains and rods have already been reported~\cite{Grains1,Rods1,Rods2,Rods3,Rods4,Rods5,Rods6,Rods7}. The topological surface states are described by the relativistic-like Dirac equation, and their dispersion is linear $\epsilon_\vec{p}=v p$ with velocity $v$. The bulk MP waves in $\hbox{Bi}_2 \hbox{Se}_3$ thin films have been previously reported~\cite{PlasmonsTI}, and we use a corresponding set of parameters. Here, we apply the Dirac velocity $v=0.67 \; 10^{6} \; \hbox{cm}^{-2}$, electron density $n\approx 7\;10^{12}~\hbox{cm}^{-2}$, and magnetic field $B\approx2\; \hbox{T}$. The corresponding Fermi energy and Larmor frequency are given by $\epsilon_\mathrm{F}\approx 0.4 \; \hbox{eV}$ and $\hbar \omega_0\approx1.2 \;\hbox{meV}$, respectively. Due to the absence of an external gate induced extra screening, the effective dielectric constant $\kappa=35$ for a $\hbox{Bi}_2 \hbox{Se}_3$-air interface can be approximated to be wave vector  independent. The screening impacts only the strength of the Coulomb interactions and their long-range nature is retained. 

\begin{figure}[t]
	\begin{center}
		\includegraphics[trim=1cm 0cm 0.15cm 0.8cm, clip, width=\columnwidth]{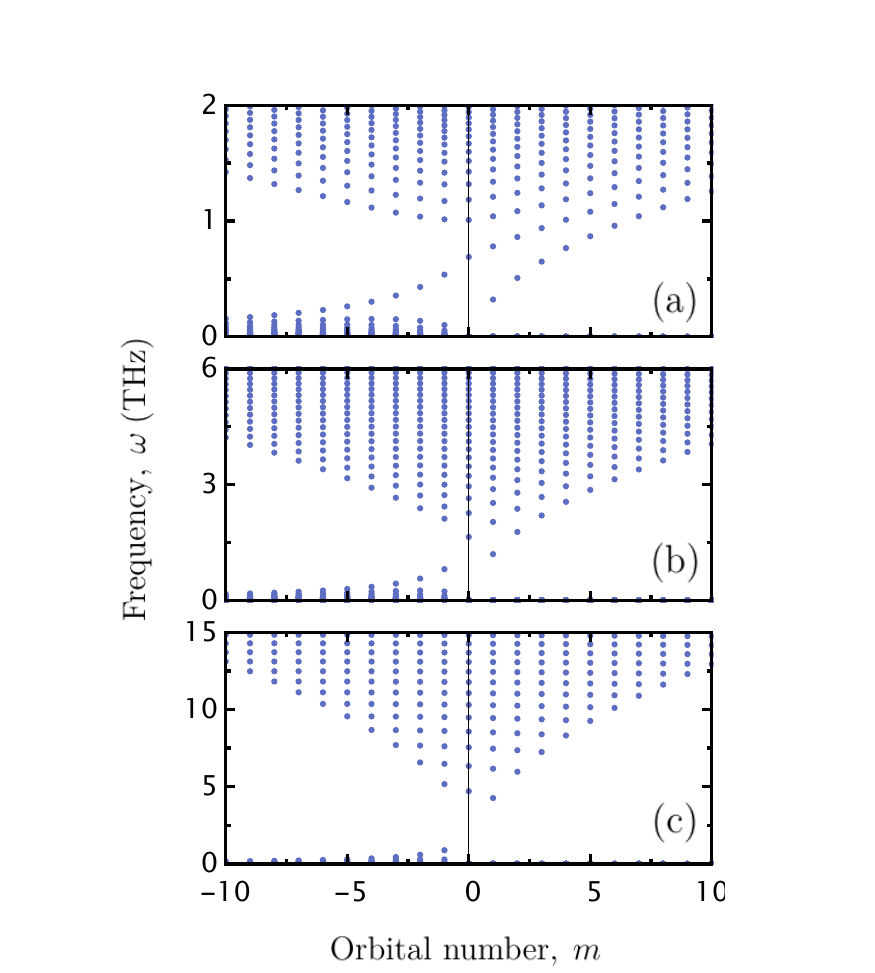}
		\caption{Spectrum of MP waves supported by a 2D electron gas confined at the surface of a sphere with radius $100\;\mu\hbox{m}$ (a), $10\;\mu\hbox{m}$ (b), or $1\;\mu\hbox{m}$ (c). The equatorial modes are well developed only if the sphere radius exceeds the penetration length $l_0\approx 5 \; \mu \hbox{m}$.}
		\label{Fig5}
	\end{center}
	\vspace{-0.2in}
\end{figure}

\begin{figure}[t]
	\begin{center}
		\includegraphics[trim=1cm 0.1cm 0.35cm 0.8cm, clip, width=\columnwidth]{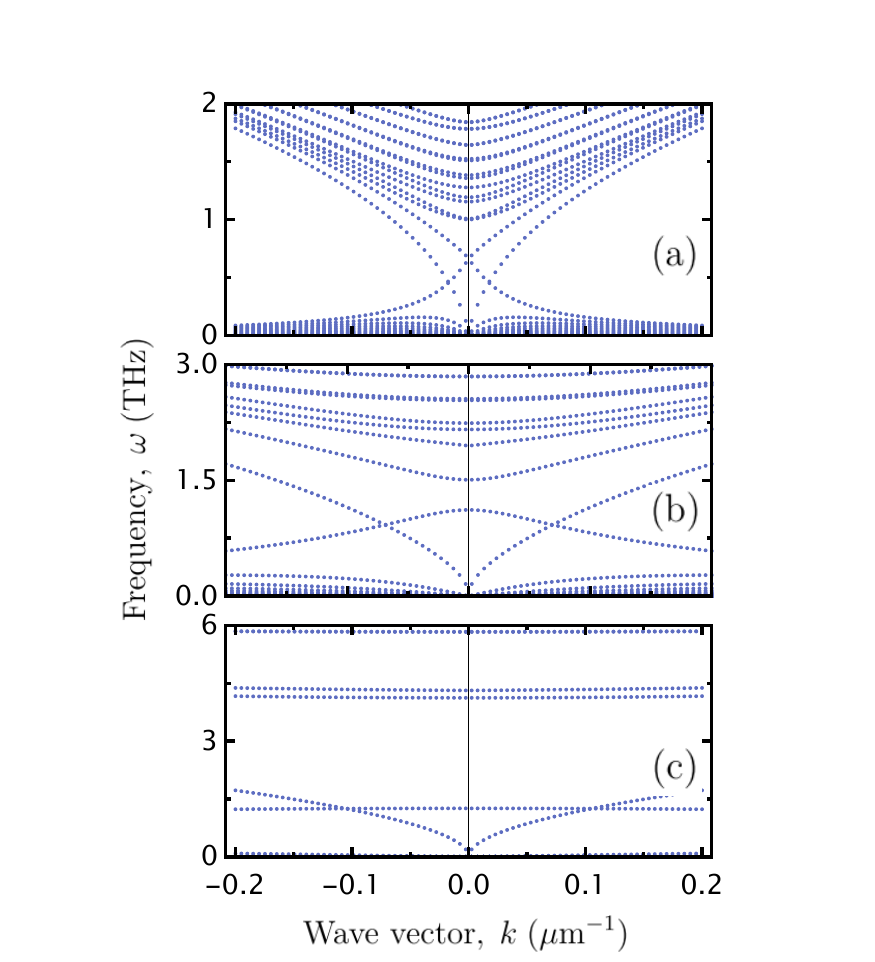}
		\caption{Spectrum of MP waves supported by a 2D electron gas confined at the surface of a cylinder with radius $100\;\mu\hbox{m}$ (a), $10\;\mu\hbox{m}$ (b), or $1\;\mu\hbox{m}$ (c). The chiral Kelvin and Yanai modes demonstrate exceptional robustness and are well developed even if the cylinder radius is smaller than the penetration length $l_0\approx 5 \; \mu \hbox{m}$. As we discuss in the main text, the robustness can be explained by destructive interference between interfacet couplings across the top and bottom hemicylinders.}
		\label{Fig5}
	\end{center}
	\vspace{-0.2in}
\end{figure}

The classical description of MP waves ignores the discrete nature of the Dirac electron spectrum due to the magnetic field and finite-size effects. The classical approach is justified by the hierarchies of energy scales, $\hbar \omega_0\ll \epsilon_\mathrm{F}$ and $\hbar \omega_\mathrm{R}^\mathrm{e}\ll \epsilon_\mathrm{F}$, where $\omega_\mathrm{R}^\mathrm{e}= v/\mathrm{R}$. These relations are well satisfied for the considered range of magnetic fields and for the $\mu$m-sized samples considered below. Another important length scale is the penetration length $l_0$ for edge MP waves. This scale can be estimated as the inverse wave vector $l_0=q^{-1}$ at which $\omega_\mathrm{P}(q)=\omega_0$. Here $\omega_\mathrm{P}(q)=\sqrt{2\pi n e^2 q\;/m \kappa}$ is dispersion of plasma waves in the absence of magnetic field and the external screening. For the considered set of parameters, we have $l_0\approx 5 ~\mu\hbox{m}$. For grains and rods with radius $R\sim l_0$, the finite-size effects become essential. It is instructive to discuss the spherical and cylindrical geometries separately.

\emph{For a spherical geometry}, the system of equations describing MP waves, Eqs.~(\ref{Continuity}) - (\ref{Potential}), can be solved via an expansion over the vector spherical harmonics. These calculations are presented in Appendix B. The spherical harmonics are labeled by the orbital discrete numbers $l=0,1,2,...$ and $m=-l, ... ,l$. In the presence of a uniform magnetic field, the axial symmetry remains, indicating that $m$ is still a good discrete number. The evaluated spectrum is presented in Fig.~3, which displays results for a sphere with radius $100\;\mu\hbox{m}$ (a), $10\;\mu\hbox{m}$ (b), or $1\;\mu\hbox{m}$ (c). For $R\gg l_0 $, the spectrum of Kelvin, Yanai, and Rossby waves is well developed, resembling the spectrum calculated for the planar geometry with overscreened interactions in Sec.~III. However, in this case, the topological Kelvin MP wave exhibits square-root behavior instead of linear behavior. The discrete nature of the spectrum can be attributed to the formation of standing waves that restrict the wave vector along the equator $k_{\mathrm{\phi}}$ as  $k_{\mathrm{\phi}} R =m$. The Poincar\'e modes become indistinguishable from the discrete modes originating from the bulk continuum. When the radius becomes comparable to the penetration length $R\lesssim l_0$, the equatorial modes are pushed outside the gap and are no longer well resolved.

\emph{For a cylindrical geometry}, the system of equations describing MP waves, Eqs.~(\ref{Continuity}) - (\ref{Potential}), can be solved via a Fourier transform and an expansion over the circular harmonics. Due to the translational symmetry along the cylinder, the spectrum can be labeled by the corresponding wave vector $k$, as presented in Fig.~4. We have used the same values for the cylinder radius: $100\;\mu\hbox{m}$ (a), $10\;\mu\hbox{m}$ (b), and $1\;\mu\hbox{m}$ (c). For $R\gg l_0 $, the spectrum represents superimposed independent spectra for two domain walls that are situated at the opposite facets of the cylinder, as illustrated in Fig.~1-b. When the radius becomes comparable to the penetration length $R\lesssim l_0$, indistinguishable discrete bulk and Poincar\'e modes are pushed outside the gap, while the Rossby modes are shifted toward larger wave vectors. The Yanai modes from different facets experience the hybridization that is the most prominent in the vicinity of their avoided crossing at $k=0$. However, this behavior is not observed for the Kelvin and Yanai waves. At first glance, their exceptional robustness and ability to overcome finite-size effects are surprising. The dispersion curves for the Kelvin and Yanai modes from opposite facets intersect. As a result, any overlap between them is expected to induce intermode hybridization and a gap between the hybrid modes. We attribute their robustness to these destructive interference effects. Because the Kelvin MP mode is longitudinal ($J^x=0$), its hybridization with the Yanai mode from the opposite facet originates from the overlap between transverse current $J^y$ and electron density $J^0$ components. As presented in Appendix A, the profiles across the domain wall are symmetric for the Kelvin mode, but antisymmetric for the Yanai mode. As a result, the overlapping regions in the top and bottom hemicylinders have opposite signs but the same magnitude; thus, their total contribution vanishes. In other words, the couplings across the top and bottom hemicylinders interfere destructively, which forbids intermode hybridization and ensures exceptional robustness of the equatorial MP mode spectrum in the cylindrical geometry. We discuss the robustness of spectra and destructive interference effects in detail in Appendix A.

\section{V. Discussion}
In our description of MP waves, we have neglected retardation effects for the electric field. These effects are of second order for the factor $\omega_0/kc$. This factor is very small in the full wave vector range, except for vanishingly small momenta, $k\lesssim\omega_0/c$~\footnote{In this regime, MP waves hybridize with light and become polaritons.}. We have also omitted the time-dependent magnetic field created by density oscillations in electron gas, which can also be treated in a self-consistent manner. This magnetic field impacts only nonlinear dynamics of MP waves. In addition, the transverse electric field generated by magnetic field oscillations is also of the second order for the small factor $\omega_0/kc$ and can be safely neglected.

The hydrodynamic description of magnetoplasma waves neglects effects of the spatial dispersion in the response of the electron gas. These effects are of little importance in a wide frequency range (including the range corresponding to equatorial magnetoplasma waves), but are essential in the vicinity of overtones, $n_\mathrm{B} \omega_0$ with integer $n_\mathrm{B}$, for the cyclotron frequency. Effects of the spatial dispersion allow the hybridization between cyclotron resonance overtones and magnetoplasma waves that results in a formation of Bernstein modes~\cite{BernsteinTheoryClassic1,BernsteinTheoryClassic2,BernsteinTheoryGraphene1,BernsteinTheoryGraphene2}. The latter have been observed as in conventional conventional GaAs/AlGaAs heterostructures~\cite{BersteinModesExp1,BersteinModesExp2,BersteinModesExp3,BersteinModesExp4} as in graphene~\cite{BersteinModesExpGraphene}.

Edge MP waves localized at sample edges (including edge states in arrays of metallic disks or ribbons) have been extensively studied in conventional GaAs/AlGaAs heterostructures~\cite{PlasmonsQW1}, graphene~\cite{PlasmonsGraphene1,PlasmonsGraphene2,PlasmonsGraphene3}, and topological insulators~\cite{PlasmonsTI}. In contrast, magnetic field domain wall MP states have been reported only very recently in a circular-shaped domain wall imprinted in a GaAs/AlGaAs heterostructure~\cite{MagnetoPlasmons2}. These states have been effectively probed via near-field radiation. However, due to the narrow frequency resolution of corresponding waveguides, only Kelvin MP waves have been reported. This approach, as well as other near-field radiation approaches, can be employed to detect equatorial MP waves.   

Due to the discrete nature of their spectrum, equatorial MP waves in the spherical geometry can also be optically probed (far-field regime). In the dipole approximation, only modes with $m=\pm1$ are optically active, which can be seen as the selection rules for MP waves. Symmetry analysis demonstrates that the equatorial modes with higher $m$ values can be excited via optical vortex beams. It has already been demonstrated that these beams can resonantly couple with quadrupole and hexapole plasma modes in metallic nanostructures (e.g., discs) \cite{OpticalVorticities1,OpticalVorticities2}, and 
we anticipate that they can be used to resonantly excite equatorial MP waves. Alternatively, equatorial MP waves can be probed via scattering of light by the grain. The corresponding theory of Mie scattering has already been extended to grains and rods made of topological insulators~\cite{MieSphere1,MieSphere1,MieSphere2,MieSphere3}, but in the absence of external an uniform magnetic field, which is an essential ingredient for the formation of equatorial MPs. The magnetic field breaks the spherical symmetry that is the cornerstone assumption of the Mie theory, and its extension is beyond the scope of the present work. 

The employed classical description of MP waves is justified by the hierarchy of energy scales, $\hbar \omega_\mathrm{R}^\mathrm{e}\ll \epsilon_\mathrm{F}$, which is well satisfied for the $\mu$m-sized samples considered in this work. It should be mentioned that nm-sized topological insulator structures have previously attracted much attention~\cite{ABCylinder1,ABCylinder2,ABSpherical1,ABSpherical2,ABSpherical3} because the curved nature of the grain or rod surface results in the spin Berry phase for Dirac electrons. The latter can be described by the effective vector potential associated with a fictitious magnetic monopole induced at the center of the grain~\cite{ABSpherical1,ABSpherical2,ABSpherical3} or a magnetic flux penetrating the rod~\cite{ABCylinder1,ABCylinder2}. Via the Aharonov-Bohm effect, these systems exhibit a shift in the single-particle spectrum for Dirac electrons. These effects cannot be captured by the classical approach, but are only important for nm-sized samples. 

Grain and rods composed of a topological insulator (e.g., $\hbox{Bi}_2 \hbox{Se}_3$ or $\hbox{Bi}_2 \hbox{Te}_3$) represent a promising setup for experimental observation of equatorial MP waves. However, the unconventional physics of Dirac surface states, which are also manifested in plasmonics~\cite{SpinPlasmonsRaghu,SpinPlasmonsEfimkin1,SpinPlasmonsEfimkin2,SpinPlasmonsStauber1,SpinPlasmonsEfimkin3,SpinPlasmonsReview1}, are of little importance. Another possible setup involves conventional metal-coated dielectric particles (e.g., $\hbox{Au}_2\hbox{S}$ ~\cite{MetalicNanoShell1,MetalicNanoShell2,MetalicNanoShell3}). While spherical and cylindrical geometries are convenient for theoretical analysis, the topological nature of equatorial MP waves guarantees their presence for any closed conducting surface penetrated by a uniform magnetic field. 

We have demonstrated that the domain wall, where the magnetic field is smooth and switches its sign, hosts four distinct MP waves, including Kelvin, Yanai, Rossby, and Poincar\'e modes. It should be noted that MP edge modes of a different physical origin can be hosted at domain walls separating regions with different electron densities~\cite{EdgeDensity1,EdgeDensity2,EdgeDensity3} or the anomalous Hall conductivities~\cite{EdgeAnomalous1,EdgeAnomalous2}, as well as at edges of anisotropic two-dimensional materials~\cite{Edge2DAnisotropic}.

To conclude, we have considered plasma oscillations supported by a 2D electron gas confined at the surface of a sphere or cylinder. We argue that in the presence of a uniform magnetic field, these systems host a set of equatorial MP waves that represent counterparts to the equatorial waves trapped by Earth due to its rotation.

\section{Acknowledgements}

We acknowledge support from the Australian Research Council Centre of Excellence in Future Low-Energy Electronics Technologies.

\bibliography{ReferencesEqMPs}

\begin{thebibliography}{88}%
\makeatletter
\providecommand \@ifxundefined [1]{%
 \@ifx{#1\undefined}
}%
\providecommand \@ifnum [1]{%
 \ifnum #1\expandafter \@firstoftwo
 \else \expandafter \@secondoftwo
 \fi
}%
\providecommand \@ifx [1]{%
 \ifx #1\expandafter \@firstoftwo
 \else \expandafter \@secondoftwo
 \fi
}%
\providecommand \natexlab [1]{#1}%
\providecommand \enquote  [1]{``#1''}%
\providecommand \bibnamefont  [1]{#1}%
\providecommand \bibfnamefont [1]{#1}%
\providecommand \citenamefont [1]{#1}%
\providecommand \href@noop [0]{\@secondoftwo}%
\providecommand \href [0]{\begingroup \@sanitize@url \@href}%
\providecommand \@href[1]{\@@startlink{#1}\@@href}%
\providecommand \@@href[1]{\endgroup#1\@@endlink}%
\providecommand \@sanitize@url [0]{\catcode `\\12\catcode `\$12\catcode
  `\&12\catcode `\#12\catcode `\^12\catcode `\_12\catcode `\%12\relax}%
\providecommand \@@startlink[1]{}%
\providecommand \@@endlink[0]{}%
\providecommand \url  [0]{\begingroup\@sanitize@url \@url }%
\providecommand \@url [1]{\endgroup\@href {#1}{\urlprefix }}%
\providecommand \urlprefix  [0]{URL }%
\providecommand \Eprint [0]{\href }%
\providecommand \doibase [0]{http://dx.doi.org/}%
\providecommand \selectlanguage [0]{\@gobble}%
\providecommand \bibinfo  [0]{\@secondoftwo}%
\providecommand \bibfield  [0]{\@secondoftwo}%
\providecommand \translation [1]{[#1]}%
\providecommand \BibitemOpen [0]{}%
\providecommand \bibitemStop [0]{}%
\providecommand \bibitemNoStop [0]{.\EOS\space}%
\providecommand \EOS [0]{\spacefactor3000\relax}%
\providecommand \BibitemShut  [1]{\csname bibitem#1\endcsname}%
\let\auto@bib@innerbib\@empty
%</preamble>
\bibitem [{\citenamefont {Burkov}(2018)}]{WeylReview}%
  \BibitemOpen
  \bibfield  {author} {\bibinfo {author} {\bibfnamefont {A.}~\bibnamefont
  {Burkov}},\ }\bibfield  {title} {\bibinfo {title} {\emph {Weyl Metals}},\
  }\href {\doibase 10.1146/annurev-conmatphys-033117-054129} {\bibfield
  {journal} {\bibinfo  {journal} {Annual Review of Condensed Matter Physics}\
  }\textbf {\bibinfo {volume} {9}},\ \bibinfo {pages} {359} (\bibinfo {year}
  {2018})}\BibitemShut {NoStop}%
\bibitem [{\citenamefont {Hasan}\ and\ \citenamefont
  {Kane}(2010)}]{TopologicalInsulatorsReview}%
  \BibitemOpen
  \bibfield  {author} {\bibinfo {author} {\bibfnamefont {M.~Z.}\ \bibnamefont
  {Hasan}}\ and\ \bibinfo {author} {\bibfnamefont {C.~L.}\ \bibnamefont
  {Kane}},\ }\bibfield  {title} {\bibinfo {title} {\emph {Colloquium:
  Topological insulators}},\ }\href {\doibase 10.1103/RevModPhys.82.3045}
  {\bibfield  {journal} {\bibinfo  {journal} {Rev. Mod. Phys.}\ }\textbf
  {\bibinfo {volume} {82}},\ \bibinfo {pages} {3045} (\bibinfo {year}
  {2010})}\BibitemShut {NoStop}%
\bibitem [{\citenamefont {Sato}\ and\ \citenamefont
  {Ando}(2017)}]{TopologicalSuperconductprsReview}%
  \BibitemOpen
  \bibfield  {author} {\bibinfo {author} {\bibfnamefont {M.}~\bibnamefont
  {Sato}}\ and\ \bibinfo {author} {\bibfnamefont {Y.}~\bibnamefont {Ando}},\
  }\bibfield  {title} {\bibinfo {title} {\emph {Topological superconductors: a
  review}},\ }\href {\doibase 10.1088/1361-6633/aa6ac7} {\bibfield  {journal}
  {\bibinfo  {journal} {Reports on Progress in Physics}\ }\textbf {\bibinfo
  {volume} {80}},\ \bibinfo {pages} {076501} (\bibinfo {year}
  {2017})}\BibitemShut {NoStop}%
\bibitem [{\citenamefont {Lu}\ \emph {et~al.}(2014)\citenamefont {Lu},
  \citenamefont {Joannopoulos},\ and\ \citenamefont
  {Solja{\v{c}}i{\'{c}}}}]{TopologicalPhotonicsReview1}%
  \BibitemOpen
  \bibfield  {author} {\bibinfo {author} {\bibfnamefont {L.}~\bibnamefont
  {Lu}}, \bibinfo {author} {\bibfnamefont {J.~D.}\ \bibnamefont
  {Joannopoulos}}, \ and\ \bibinfo {author} {\bibfnamefont {M.}~\bibnamefont
  {Solja{\v{c}}i{\'{c}}}},\ }\bibfield  {title} {\bibinfo {title} {\emph
  {Topological photonics}},\ }\href {\doibase 10.1038/nphoton.2014.248}
  {\bibfield  {journal} {\bibinfo  {journal} {Nature Photonics}\ }\textbf
  {\bibinfo {volume} {8}},\ \bibinfo {pages} {821} (\bibinfo {year}
  {2014})}\BibitemShut {NoStop}%
\bibitem [{\citenamefont {Ozawa}\ \emph {et~al.}(2019)\citenamefont {Ozawa},
  \citenamefont {Price}, \citenamefont {Amo}, \citenamefont {Goldman},
  \citenamefont {Hafezi}, \citenamefont {Lu}, \citenamefont {Rechtsman},
  \citenamefont {Schuster}, \citenamefont {Simon}, \citenamefont {Zilberberg},\
  and\ \citenamefont {Carusotto}}]{TopologicalPhotonicsReview2}%
  \BibitemOpen
  \bibfield  {author} {\bibinfo {author} {\bibfnamefont {T.}~\bibnamefont
  {Ozawa}}, \bibinfo {author} {\bibfnamefont {H.~M.}\ \bibnamefont {Price}},
  \bibinfo {author} {\bibfnamefont {A.}~\bibnamefont {Amo}}, \bibinfo {author}
  {\bibfnamefont {N.}~\bibnamefont {Goldman}}, \bibinfo {author} {\bibfnamefont
  {M.}~\bibnamefont {Hafezi}}, \bibinfo {author} {\bibfnamefont
  {L.}~\bibnamefont {Lu}}, \bibinfo {author} {\bibfnamefont {M.~C.}\
  \bibnamefont {Rechtsman}}, \bibinfo {author} {\bibfnamefont {D.}~\bibnamefont
  {Schuster}}, \bibinfo {author} {\bibfnamefont {J.}~\bibnamefont {Simon}},
  \bibinfo {author} {\bibfnamefont {O.}~\bibnamefont {Zilberberg}}, \ and\
  \bibinfo {author} {\bibfnamefont {I.}~\bibnamefont {Carusotto}},\ }\bibfield
  {title} {\bibinfo {title} {\emph {Topological photonics}},\ }\href {\doibase
  10.1103/RevModPhys.91.015006} {\bibfield  {journal} {\bibinfo  {journal}
  {Rev. Mod. Phys.}\ }\textbf {\bibinfo {volume} {91}},\ \bibinfo {pages}
  {015006} (\bibinfo {year} {2019})}\BibitemShut {NoStop}%
\bibitem [{\citenamefont {Kim}\ \emph {et~al.}(2020)\citenamefont {Kim},
  \citenamefont {Jacob},\ and\ \citenamefont
  {Rho}}]{TopologicalPhotonicsReview3}%
  \BibitemOpen
  \bibfield  {author} {\bibinfo {author} {\bibfnamefont {M.}~\bibnamefont
  {Kim}}, \bibinfo {author} {\bibfnamefont {Z.}~\bibnamefont {Jacob}}, \ and\
  \bibinfo {author} {\bibfnamefont {J.}~\bibnamefont {Rho}},\ }\bibfield
  {title} {\bibinfo {title} {\emph {Recent advances in 2D, 3D and higher-order
  topological photonics}},\ }\href {\doibase 10.1038/s41377-020-0331-y}
  {\bibfield  {journal} {\bibinfo  {journal} {Light: Science {\&}
  Applications}\ }\textbf {\bibinfo {volume} {9}},\ \bibinfo {pages} {130}
  (\bibinfo {year} {2020})}\BibitemShut {NoStop}%
\bibitem [{\citenamefont {Lee}\ \emph {et~al.}(2018)\citenamefont {Lee},
  \citenamefont {Imhof}, \citenamefont {Berger}, \citenamefont {Bayer},
  \citenamefont {Brehm}, \citenamefont {Molenkamp}, \citenamefont {Kiessling},\
  and\ \citenamefont {Thomale}}]{TopologicalCircuits1}%
  \BibitemOpen
  \bibfield  {author} {\bibinfo {author} {\bibfnamefont {C.~H.}\ \bibnamefont
  {Lee}}, \bibinfo {author} {\bibfnamefont {S.}~\bibnamefont {Imhof}}, \bibinfo
  {author} {\bibfnamefont {C.}~\bibnamefont {Berger}}, \bibinfo {author}
  {\bibfnamefont {F.}~\bibnamefont {Bayer}}, \bibinfo {author} {\bibfnamefont
  {J.}~\bibnamefont {Brehm}}, \bibinfo {author} {\bibfnamefont {L.~W.}\
  \bibnamefont {Molenkamp}}, \bibinfo {author} {\bibfnamefont {T.}~\bibnamefont
  {Kiessling}}, \ and\ \bibinfo {author} {\bibfnamefont {R.}~\bibnamefont
  {Thomale}},\ }\bibfield  {title} {\bibinfo {title} {\emph {Topolectrical
  Circuits}},\ }\href {\doibase 10.1038/s42005-018-0035-2} {\bibfield
  {journal} {\bibinfo  {journal} {Communications Physics}\ }\textbf {\bibinfo
  {volume} {1}},\ \bibinfo {pages} {39} (\bibinfo {year} {2018})}\BibitemShut
  {NoStop}%
\bibitem [{\citenamefont {Dong}\ \emph {et~al.}(2021)\citenamefont {Dong},
  \citenamefont {Juri\ifmmode \check{c}\else \v{c}\fi{}i\ifmmode~\acute{c}\else
  \'{c}\fi{}},\ and\ \citenamefont {Roy}}]{TopologicalCircuits2}%
  \BibitemOpen
  \bibfield  {author} {\bibinfo {author} {\bibfnamefont {J.}~\bibnamefont
  {Dong}}, \bibinfo {author} {\bibfnamefont {V.}~\bibnamefont {Juri\ifmmode
  \check{c}\else \v{c}\fi{}i\ifmmode~\acute{c}\else \'{c}\fi{}}}, \ and\
  \bibinfo {author} {\bibfnamefont {B.}~\bibnamefont {Roy}},\ }\bibfield
  {title} {\bibinfo {title} {\emph {Topolectric circuits: Theory and
  construction}},\ }\href {\doibase 10.1103/PhysRevResearch.3.023056}
  {\bibfield  {journal} {\bibinfo  {journal} {Phys. Rev. Research}\ }\textbf
  {\bibinfo {volume} {3}},\ \bibinfo {pages} {023056} (\bibinfo {year}
  {2021})}\BibitemShut {NoStop}%
\bibitem [{\citenamefont {Bergholtz}\ \emph {et~al.}(2021)\citenamefont
  {Bergholtz}, \citenamefont {Budich},\ and\ \citenamefont
  {Kunst}}]{NonHermitianTopology1}%
  \BibitemOpen
  \bibfield  {author} {\bibinfo {author} {\bibfnamefont {E.~J.}\ \bibnamefont
  {Bergholtz}}, \bibinfo {author} {\bibfnamefont {J.~C.}\ \bibnamefont
  {Budich}}, \ and\ \bibinfo {author} {\bibfnamefont {F.~K.}\ \bibnamefont
  {Kunst}},\ }\bibfield  {title} {\bibinfo {title} {\emph {Exceptional topology
  of non-Hermitian systems}},\ }\href {\doibase 10.1103/RevModPhys.93.015005}
  {\bibfield  {journal} {\bibinfo  {journal} {Rev. Mod. Phys.}\ }\textbf
  {\bibinfo {volume} {93}},\ \bibinfo {pages} {015005} (\bibinfo {year}
  {2021})}\BibitemShut {NoStop}%
\bibitem [{\citenamefont {Kawabata}\ \emph {et~al.}(2019)\citenamefont
  {Kawabata}, \citenamefont {Shiozaki}, \citenamefont {Ueda},\ and\
  \citenamefont {Sato}}]{NonHermitianTopology2}%
  \BibitemOpen
  \bibfield  {author} {\bibinfo {author} {\bibfnamefont {K.}~\bibnamefont
  {Kawabata}}, \bibinfo {author} {\bibfnamefont {K.}~\bibnamefont {Shiozaki}},
  \bibinfo {author} {\bibfnamefont {M.}~\bibnamefont {Ueda}}, \ and\ \bibinfo
  {author} {\bibfnamefont {M.}~\bibnamefont {Sato}},\ }\bibfield  {title}
  {\bibinfo {title} {\emph {Symmetry and Topology in Non-Hermitian Physics}},\
  }\href {\doibase 10.1103/PhysRevX.9.041015} {\bibfield  {journal} {\bibinfo
  {journal} {Phys. Rev. X}\ }\textbf {\bibinfo {volume} {9}},\ \bibinfo {pages}
  {041015} (\bibinfo {year} {2019})}\BibitemShut {NoStop}%
\bibitem [{\citenamefont {Yoshida}\ and\ \citenamefont
  {Hatsugai}(2019)}]{NHadditional}%
  \BibitemOpen
  \bibfield  {author} {\bibinfo {author} {\bibfnamefont {T.}~\bibnamefont
  {Yoshida}}\ and\ \bibinfo {author} {\bibfnamefont {Y.}~\bibnamefont
  {Hatsugai}},\ }\bibfield  {title} {\bibinfo {title} {\emph {Exceptional rings
  protected by emergent symmetry for mechanical systems}},\ }\href {\doibase
  10.1103/PhysRevB.100.054109} {\bibfield  {journal} {\bibinfo  {journal}
  {Phys. Rev. B}\ }\textbf {\bibinfo {volume} {100}},\ \bibinfo {pages}
  {054109} (\bibinfo {year} {2019})}\BibitemShut {NoStop}%
\bibitem [{\citenamefont {Matsuno}(1966)}]{EquatorialClasssics}%
  \BibitemOpen
  \bibfield  {author} {\bibinfo {author} {\bibfnamefont {T.}~\bibnamefont
  {Matsuno}},\ }\bibfield  {title} {\bibinfo {title} {\emph {Quasi-Geostrophic
  Motions in the Equatorial Area}},\ }\href {\doibase 10.2151/jmsj1965.44.1_25}
  {\bibfield  {journal} {\bibinfo  {journal} {Journal of the Meteorological
  Society of Japan. Ser. II}\ }\textbf {\bibinfo {volume} {44}},\ \bibinfo
  {pages} {25} (\bibinfo {year} {1966})}\BibitemShut {NoStop}%
\bibitem [{\citenamefont {Vallis}(2006)}]{BookVallis06}%
  \BibitemOpen
  \bibfield  {author} {\bibinfo {author} {\bibfnamefont {G.~K.}\ \bibnamefont
  {Vallis}},\ }\href@noop {} {\emph {\bibinfo {title} {Atmospheric and Oceanic
  Fluid Dynamics}}}\ (\bibinfo  {publisher} {Cambridge University Press},\
  \bibinfo {address} {Cambridge, U.K.},\ \bibinfo {year} {2006})\ p.\ \bibinfo
  {pages} {745}\BibitemShut {NoStop}%
\bibitem [{\citenamefont {Delplace}\ \emph {et~al.}(2017)\citenamefont
  {Delplace}, \citenamefont {Marston},\ and\ \citenamefont
  {Venaille}}]{TopologyEquatorialWaves}%
  \BibitemOpen
  \bibfield  {author} {\bibinfo {author} {\bibfnamefont {P.}~\bibnamefont
  {Delplace}}, \bibinfo {author} {\bibfnamefont {J.~B.}\ \bibnamefont
  {Marston}}, \ and\ \bibinfo {author} {\bibfnamefont {A.}~\bibnamefont
  {Venaille}},\ }\bibfield  {title} {\bibinfo {title} {\emph {Topological
  origin of equatorial waves}},\ }\href {\doibase 10.1126/science.aan8819}
  {\bibfield  {journal} {\bibinfo  {journal} {Science}\ }\textbf {\bibinfo
  {volume} {358}},\ \bibinfo {pages} {1075} (\bibinfo {year}
  {2017})}\BibitemShut {NoStop}%
\bibitem [{\citenamefont {Dyakonov}\ and\ \citenamefont
  {Shur}(1993)}]{DiakonovShur}%
  \BibitemOpen
  \bibfield  {author} {\bibinfo {author} {\bibfnamefont {M.}~\bibnamefont
  {Dyakonov}}\ and\ \bibinfo {author} {\bibfnamefont {M.}~\bibnamefont
  {Shur}},\ }\bibfield  {title} {\bibinfo {title} {\emph {Shallow water analogy
  for a ballistic field effect transistor: New mechanism of plasma wave
  generation by dc current}},\ }\href {\doibase 10.1103/PhysRevLett.71.2465}
  {\bibfield  {journal} {\bibinfo  {journal} {Phys. Rev. Lett.}\ }\textbf
  {\bibinfo {volume} {71}},\ \bibinfo {pages} {2465} (\bibinfo {year}
  {1993})}\BibitemShut {NoStop}%
\bibitem [{\citenamefont {Chaplik}(1972)}]{Chaplik1972PossibleCO}%
  \BibitemOpen
  \bibfield  {author} {\bibinfo {author} {\bibfnamefont {A.~V.}\ \bibnamefont
  {Chaplik}},\ }\bibfield  {title} {\bibinfo {title} {\emph {Possible
  Crystallization of Charge Carriers in Low-density Inversion Layers}},\
  }\href@noop {} {\bibfield  {journal} {\bibinfo  {journal} {Journal of
  Experimental and Theoretical Physics}\ }\textbf {\bibinfo {volume} {35}},\
  \bibinfo {pages} {395} (\bibinfo {year} {1972})}\BibitemShut {NoStop}%
\bibitem [{Note1()}]{Note1}%
  \BibitemOpen
  \bibinfo {note} {Plasma waves in the presence of a magnetic field are usually
  referred to as magnetoplasma waves or magnetoplasmons}\BibitemShut {NoStop}%
\bibitem [{\citenamefont {Jin}\ \emph {et~al.}(2016)\citenamefont {Jin},
  \citenamefont {Lu}, \citenamefont {Wang}, \citenamefont {Fang}, \citenamefont
  {Joannopoulos}, \citenamefont {Solja{\v{c}}i{\'{c}}}, \citenamefont {Fu},\
  and\ \citenamefont {Fang}}]{MagnetoPlasmons1}%
  \BibitemOpen
  \bibfield  {author} {\bibinfo {author} {\bibfnamefont {D.}~\bibnamefont
  {Jin}}, \bibinfo {author} {\bibfnamefont {L.}~\bibnamefont {Lu}}, \bibinfo
  {author} {\bibfnamefont {Z.}~\bibnamefont {Wang}}, \bibinfo {author}
  {\bibfnamefont {C.}~\bibnamefont {Fang}}, \bibinfo {author} {\bibfnamefont
  {J.~D.}\ \bibnamefont {Joannopoulos}}, \bibinfo {author} {\bibfnamefont
  {M.}~\bibnamefont {Solja{\v{c}}i{\'{c}}}}, \bibinfo {author} {\bibfnamefont
  {L.}~\bibnamefont {Fu}}, \ and\ \bibinfo {author} {\bibfnamefont {N.~X.}\
  \bibnamefont {Fang}},\ }\bibfield  {title} {\bibinfo {title} {\emph
  {Topological magnetoplasmon}},\ }\href {\doibase 10.1038/ncomms13486}
  {\bibfield  {journal} {\bibinfo  {journal} {Nature Communications}\ }\textbf
  {\bibinfo {volume} {7}},\ \bibinfo {pages} {13486} (\bibinfo {year}
  {2016})}\BibitemShut {NoStop}%
\bibitem [{\citenamefont {Fetter}(1985)}]{Fetter1}%
  \BibitemOpen
  \bibfield  {author} {\bibinfo {author} {\bibfnamefont {A.~L.}\ \bibnamefont
  {Fetter}},\ }\bibfield  {title} {\bibinfo {title} {\emph {Edge
  magnetoplasmons in a bounded two-dimensional electron fluid}},\ }\href
  {\doibase 10.1103/PhysRevB.32.7676} {\bibfield  {journal} {\bibinfo
  {journal} {Phys. Rev. B}\ }\textbf {\bibinfo {volume} {32}},\ \bibinfo
  {pages} {7676} (\bibinfo {year} {1985})}\BibitemShut {NoStop}%
\bibitem [{Note2()}]{Note2}%
  \BibitemOpen
  \bibinfo {note} {For electrons with conventional quadratic dispersion, this
  term is equal to their band mass. For electrons with unconventional but
  isotropic dispersion (e.g., with relativistic-like Dirac dispersion), the
  mass $m=\epsilon _\protect \mathrm {F}/v_\protect \mathrm {F}$ is given by
  the ratio between the Fermi energy $\epsilon _\protect \mathrm {F}$ and Fermi
  velocity $v_\protect \mathrm {F}$.}\BibitemShut {Stop}%
\bibitem [{Note3()}]{Note3}%
  \BibitemOpen
  \bibinfo {note} {The explicit expression for $C$ is given by $$C= \begin
  {pmatrix} 0 & 0 & \protect \mathcal {K} \\ 0 & \protect \mathcal {K} & 0 \\
  \protect \mathcal {K} & 0 & 0 \\ \end {pmatrix}, $$ where $\protect \mathcal
  {K}$ is the complex conjugation operator.}\BibitemShut {Stop}%
\bibitem [{\citenamefont {Ryu}\ \emph {et~al.}(2010)\citenamefont {Ryu},
  \citenamefont {Schnyder}, \citenamefont {Furusaki},\ and\ \citenamefont
  {Ludwig}}]{TopClassification}%
  \BibitemOpen
  \bibfield  {author} {\bibinfo {author} {\bibfnamefont {S.}~\bibnamefont
  {Ryu}}, \bibinfo {author} {\bibfnamefont {A.~P.}\ \bibnamefont {Schnyder}},
  \bibinfo {author} {\bibfnamefont {A.}~\bibnamefont {Furusaki}}, \ and\
  \bibinfo {author} {\bibfnamefont {A.~W.~W.}\ \bibnamefont {Ludwig}},\
  }\bibfield  {title} {\bibinfo {title} {\emph {Topological insulators and
  superconductors: tenfold way and dimensional hierarchy}},\ }\href {\doibase
  10.1088/1367-2630/12/6/065010} {\bibfield  {journal} {\bibinfo  {journal}
  {New Journal of Physics}\ }\textbf {\bibinfo {volume} {12}},\ \bibinfo
  {pages} {065010} (\bibinfo {year} {2010})}\BibitemShut {NoStop}%
\bibitem [{Note4()}]{Note4}%
  \BibitemOpen
  \bibinfo {note} {Explicit expressions for the components of spin-1 Pauli
  matrices $\protect \bm {\sigma }$ are given by \par \begin {equation*}\begin
  {split} \sigma _x= \begin {pmatrix} 0 & \protect \frac {1}{\protect \sqrt
  {2}} & 0 \\ \protect \frac {1}{\protect \sqrt {2}} & 0 & \protect \frac
  {1}{\protect \sqrt {2}} \\ 0 & \protect \frac {1}{\protect \sqrt {2}} & 0 \\
  \end {pmatrix}, \hskip 1em\relax \hskip 1em\relax \sigma _y= \begin {pmatrix}
  0 & - \protect \frac {i}{\protect \sqrt {2}} & 0 \\ \protect \frac
  {i}{\protect \sqrt {2}} & 0 & - \protect \frac {i}{\protect \sqrt {2}} \\ 0 &
  \protect \frac {i}{\protect \sqrt {2}} & 0 \\ \end {pmatrix}, \\ \sigma _z=
  \begin {pmatrix} 1 & 0 & 0 \\ 0 & 0 & 0 \\ 0 & 0 & -1 \\ \end
  {pmatrix}.\hskip 1em\relax \hskip 1em\relax \hskip 1em\relax \hskip 1em\relax
  \hskip 1em\relax \hskip 1em\relax \hskip 1em\relax \hskip 1em\relax \end
  {split} \end {equation*}}\BibitemShut {NoStop}%
\bibitem [{Note5()}]{Note5}%
  \BibitemOpen
  \bibinfo {note} {The motion of the electrons is confined to the plane, but
  interactions are mediated by an electric field that is extended in
  three-dimensional space.}\BibitemShut {Stop}%
\bibitem [{\citenamefont {Tauber}\ \emph {et~al.}(2020)\citenamefont {Tauber},
  \citenamefont {Delplace},\ and\ \citenamefont {Venaille}}]{ABEC1}%
  \BibitemOpen
  \bibfield  {author} {\bibinfo {author} {\bibfnamefont {C.}~\bibnamefont
  {Tauber}}, \bibinfo {author} {\bibfnamefont {P.}~\bibnamefont {Delplace}}, \
  and\ \bibinfo {author} {\bibfnamefont {A.}~\bibnamefont {Venaille}},\
  }\bibfield  {title} {\bibinfo {title} {\emph {Anomalous bulk-edge
  correspondence in continuous media}},\ }\href {\doibase
  10.1103/PhysRevResearch.2.013147} {\bibfield  {journal} {\bibinfo  {journal}
  {Phys. Rev. Research}\ }\textbf {\bibinfo {volume} {2}},\ \bibinfo {pages}
  {013147} (\bibinfo {year} {2020})}\BibitemShut {NoStop}%
\bibitem [{\citenamefont {Graf}\ \emph {et~al.}(2021)\citenamefont {Graf},
  \citenamefont {Jud},\ and\ \citenamefont {Tauber}}]{ABEC2}%
  \BibitemOpen
  \bibfield  {author} {\bibinfo {author} {\bibfnamefont {G.~M.}\ \bibnamefont
  {Graf}}, \bibinfo {author} {\bibfnamefont {H.}~\bibnamefont {Jud}}, \ and\
  \bibinfo {author} {\bibfnamefont {C.}~\bibnamefont {Tauber}},\ }\bibfield
  {title} {\bibinfo {title} {\emph {Topology in Shallow-Water Waves: A
  Violation of Bulk-Edge Correspondence}},\ }\href {\doibase
  10.1007/s00220-021-03982-7} {\bibfield  {journal} {\bibinfo  {journal}
  {Communications in Mathematical Physics}\ }\textbf {\bibinfo {volume}
  {383}},\ \bibinfo {pages} {731} (\bibinfo {year} {2021})}\BibitemShut
  {NoStop}%
\bibitem [{\citenamefont {Tauber}\ and\ \citenamefont {Thiang}(2021)}]{ABEC3}%
  \BibitemOpen
  \bibfield  {author} {\bibinfo {author} {\bibfnamefont {C.}~\bibnamefont
  {Tauber}}\ and\ \bibinfo {author} {\bibfnamefont {G.~C.}\ \bibnamefont
  {Thiang}},\ }\href@noop {} {\bibinfo {title} {\emph {Topology in
  shallow-water waves: A spectral flow perspective}}} (\bibinfo {year}
  {2021}),\ \Eprint {http://arxiv.org/abs/2110.04097} {arXiv:2110.04097
  [math-ph]} \BibitemShut {NoStop}%
\bibitem [{\citenamefont {Faure}(2019)}]{ABEC4}%
  \BibitemOpen
  \bibfield  {author} {\bibinfo {author} {\bibfnamefont {F.}~\bibnamefont
  {Faure}},\ }\href@noop {} {\bibinfo {title} {\emph {Manifestation of the
  topological index formula in quantum waves and geophysical waves}}} (\bibinfo
  {year} {2019}),\ \Eprint {http://arxiv.org/abs/1901.10592} {arXiv:1901.10592
  [math-ph]} \BibitemShut {NoStop}%
\bibitem [{\citenamefont {Tauber}\ \emph {et~al.}(2019)\citenamefont {Tauber},
  \citenamefont {Delplace},\ and\ \citenamefont {Venaille}}]{ABEC5}%
  \BibitemOpen
  \bibfield  {author} {\bibinfo {author} {\bibfnamefont {C.}~\bibnamefont
  {Tauber}}, \bibinfo {author} {\bibfnamefont {P.}~\bibnamefont {Delplace}}, \
  and\ \bibinfo {author} {\bibfnamefont {A.}~\bibnamefont {Venaille}},\
  }\bibfield  {title} {\bibinfo {title} {\emph {A bulk-interface correspondence
  for equatorial waves}},\ }\href {\doibase 10.1017/jfm.2019.233} {\bibfield
  {journal} {\bibinfo  {journal} {Journal of Fluid Mechanics}\ }\textbf
  {\bibinfo {volume} {868}},\ \bibinfo {pages} {R2} (\bibinfo {year}
  {2019})}\BibitemShut {NoStop}%
\bibitem [{\citenamefont {Fu}\ and\ \citenamefont
  {Qin}(2021)}]{BulkEdgeAdditional}%
  \BibitemOpen
  \bibfield  {author} {\bibinfo {author} {\bibfnamefont {Y.}~\bibnamefont
  {Fu}}\ and\ \bibinfo {author} {\bibfnamefont {H.}~\bibnamefont {Qin}},\
  }\bibfield  {title} {\bibinfo {title} {\emph {Topological phases and
  bulk-edge correspondence of magnetized cold plasmas}},\ }\href {\doibase
  10.1038/s41467-021-24189-3} {\bibfield  {journal} {\bibinfo  {journal}
  {Nature Communications}\ }\textbf {\bibinfo {volume} {12}},\ \bibinfo {pages}
  {3924} (\bibinfo {year} {2021})}\BibitemShut {NoStop}%
\bibitem [{Note6()}]{Note6}%
  \BibitemOpen
  \bibinfo {note} {The kink profile implies that $b(-\infty )<0$ and $b(\infty
  )>0$. It should also be noted that the mathematical structure of this
  solution is reminiscent of the celebrated Jackiw-Rebbi solution for the
  massive Dirac model with position-dependent mass~\cite
  {JackiwRebbi}.}\BibitemShut {Stop}%
\bibitem [{\citenamefont {Cooper}\ \emph {et~al.}(1995)\citenamefont {Cooper},
  \citenamefont {Khare},\ and\ \citenamefont {Sukhatme}}]{Susy1}%
  \BibitemOpen
  \bibfield  {author} {\bibinfo {author} {\bibfnamefont {F.}~\bibnamefont
  {Cooper}}, \bibinfo {author} {\bibfnamefont {A.}~\bibnamefont {Khare}}, \
  and\ \bibinfo {author} {\bibfnamefont {U.}~\bibnamefont {Sukhatme}},\
  }\bibfield  {title} {\bibinfo {title} {\emph {Supersymmetry and quantum
  mechanics}},\ }\href
  {https://www.sciencedirect.com/science/article/pii/037015739400080M}
  {\bibfield  {journal} {\bibinfo  {journal} {Physics Reports}\ }\textbf
  {\bibinfo {volume} {251}},\ \bibinfo {pages} {267} (\bibinfo {year}
  {1995})}\BibitemShut {NoStop}%
\bibitem [{\citenamefont {{\c{C}}evik}\ \emph {et~al.}(2016)\citenamefont
  {{\c{C}}evik}, \citenamefont {Gadella}, \citenamefont {Kuru},\ and\
  \citenamefont {Negro}}]{Susy2}%
  \BibitemOpen
  \bibfield  {author} {\bibinfo {author} {\bibfnamefont {D.}~\bibnamefont
  {{\c{C}}evik}}, \bibinfo {author} {\bibfnamefont {M.}~\bibnamefont
  {Gadella}}, \bibinfo {author} {\bibfnamefont {{\c{S}}.}~\bibnamefont {Kuru}},
  \ and\ \bibinfo {author} {\bibfnamefont {J.}~\bibnamefont {Negro}},\
  }\bibfield  {title} {\bibinfo {title} {\emph {Resonances and antibound states
  for the P{\"o}schl--Teller potential: Ladder operators and SUSY partners}},\
  }\href {https://www.sciencedirect.com/science/article/pii/S0375960116002255}
  {\bibfield  {journal} {\bibinfo  {journal} {Physics Letters A}\ }\textbf
  {\bibinfo {volume} {380}},\ \bibinfo {pages} {1600} (\bibinfo {year}
  {2016})}\BibitemShut {NoStop}%
\bibitem [{Note7()}]{Note7}%
  \BibitemOpen
  \bibinfo {note} {At $k=0$, the prefactor $\lambda (\lambda +1)=\omega
  _0^2/\omega _\protect \mathrm {d}^2$ is positive, which ensures the presence
  of at least one bound state trapped by the PT quantum well
  potential.}\BibitemShut {Stop}%
\bibitem [{Note8()}]{Note8}%
  \BibitemOpen
  \bibinfo {note} {We use $\protect \floor {...}$ for the integer part of a
  real number.}\BibitemShut {Stop}%
\bibitem [{Note9()}]{Note9}%
  \BibitemOpen
  \bibinfo {note} {In the presence of the domain wall, the inversion symmetry
  is broken, and these modes are no longer pinned at zero
  frequency.}\BibitemShut {Stop}%
\bibitem [{\citenamefont {Zaqarashvili}\ \emph {et~al.}(2021)\citenamefont
  {Zaqarashvili}, \citenamefont {Albekioni}, \citenamefont {Ballester},
  \citenamefont {Bekki}, \citenamefont {Biancofiore}, \citenamefont {Birch},
  \citenamefont {Dikpati}, \citenamefont {Gizon}, \citenamefont
  {Gurgenashvili}, \citenamefont {Heifetz}, \citenamefont {Lanza},
  \citenamefont {McIntosh}, \citenamefont {Ofman}, \citenamefont {Oliver},
  \citenamefont {Proxauf}, \citenamefont {Umurhan},\ and\ \citenamefont
  {Yellin-Bergovoy}}]{RossbyReview}%
  \BibitemOpen
  \bibfield  {author} {\bibinfo {author} {\bibfnamefont {T.~V.}\ \bibnamefont
  {Zaqarashvili}}, \bibinfo {author} {\bibfnamefont {M.}~\bibnamefont
  {Albekioni}}, \bibinfo {author} {\bibfnamefont {J.~L.}\ \bibnamefont
  {Ballester}}, \bibinfo {author} {\bibfnamefont {Y.}~\bibnamefont {Bekki}},
  \bibinfo {author} {\bibfnamefont {L.}~\bibnamefont {Biancofiore}}, \bibinfo
  {author} {\bibfnamefont {A.~C.}\ \bibnamefont {Birch}}, \bibinfo {author}
  {\bibfnamefont {M.}~\bibnamefont {Dikpati}}, \bibinfo {author} {\bibfnamefont
  {L.}~\bibnamefont {Gizon}}, \bibinfo {author} {\bibfnamefont
  {E.}~\bibnamefont {Gurgenashvili}}, \bibinfo {author} {\bibfnamefont
  {E.}~\bibnamefont {Heifetz}}, \bibinfo {author} {\bibfnamefont {A.~F.}\
  \bibnamefont {Lanza}}, \bibinfo {author} {\bibfnamefont {S.~W.}\ \bibnamefont
  {McIntosh}}, \bibinfo {author} {\bibfnamefont {L.}~\bibnamefont {Ofman}},
  \bibinfo {author} {\bibfnamefont {R.}~\bibnamefont {Oliver}}, \bibinfo
  {author} {\bibfnamefont {B.}~\bibnamefont {Proxauf}}, \bibinfo {author}
  {\bibfnamefont {O.~M.}\ \bibnamefont {Umurhan}}, \ and\ \bibinfo {author}
  {\bibfnamefont {R.}~\bibnamefont {Yellin-Bergovoy}},\ }\bibfield  {title}
  {\bibinfo {title} {\emph {Rossby Waves in Astrophysics}},\ }\href {\doibase
  10.1007/s11214-021-00790-2} {\bibfield  {journal} {\bibinfo  {journal} {Space
  Science Reviews}\ }\textbf {\bibinfo {volume} {217}},\ \bibinfo {pages} {15}
  (\bibinfo {year} {2021})}\BibitemShut {NoStop}%
\bibitem [{Note10()}]{Note10}%
  \BibitemOpen
  \bibinfo {note} {We ignore the disorder-induced residual metallic bulk
  conductivity that usually appears in topological insulators, including $\hbox
  {Bi}_2 \hbox {Se}_3$ samples. For thin films with thickness $l\lesssim 0.3
  \protect \tmspace +\thickmuskip {.2777em} \mu \hbox {m}$, the response has
  been argued to be fully dominated by the Dirac states.}\BibitemShut {Stop}%
\bibitem [{\citenamefont {Campos}\ \emph {et~al.}(2018)\citenamefont {Campos},
  \citenamefont {Fonseca}, \citenamefont {de~Carvalho}, \citenamefont {Mendes},
  \citenamefont {Rocha},\ and\ \citenamefont {Moura-Melo}}]{Grains1}%
  \BibitemOpen
  \bibfield  {author} {\bibinfo {author} {\bibfnamefont {W.~H.}\ \bibnamefont
  {Campos}}, \bibinfo {author} {\bibfnamefont {J.~M.}\ \bibnamefont {Fonseca}},
  \bibinfo {author} {\bibfnamefont {V.~E.}\ \bibnamefont {de~Carvalho}},
  \bibinfo {author} {\bibfnamefont {J.~B.~S.}\ \bibnamefont {Mendes}}, \bibinfo
  {author} {\bibfnamefont {M.~S.}\ \bibnamefont {Rocha}}, \ and\ \bibinfo
  {author} {\bibfnamefont {W.~A.}\ \bibnamefont {Moura-Melo}},\ }\bibfield
  {title} {\bibinfo {title} {\emph {Topological Insulator Particles As
  Optically Induced Oscillators: Toward Dynamical Force Measurements and
  Optical Rheology}},\ }\href {\doibase 10.1021/acsphotonics.7b01322}
  {\bibfield  {journal} {\bibinfo  {journal} {ACS Photonics}\ }\textbf
  {\bibinfo {volume} {5}},\ \bibinfo {pages} {741} (\bibinfo {year}
  {2018})}\BibitemShut {NoStop}%
\bibitem [{\citenamefont {Hamdou}\ \emph {et~al.}(2013)\citenamefont {Hamdou},
  \citenamefont {Gooth}, \citenamefont {Dorn}, \citenamefont {Pippel},\ and\
  \citenamefont {Nielsch}}]{Rods1}%
  \BibitemOpen
  \bibfield  {author} {\bibinfo {author} {\bibfnamefont {B.}~\bibnamefont
  {Hamdou}}, \bibinfo {author} {\bibfnamefont {J.}~\bibnamefont {Gooth}},
  \bibinfo {author} {\bibfnamefont {A.}~\bibnamefont {Dorn}}, \bibinfo {author}
  {\bibfnamefont {E.}~\bibnamefont {Pippel}}, \ and\ \bibinfo {author}
  {\bibfnamefont {K.}~\bibnamefont {Nielsch}},\ }\bibfield  {title} {\bibinfo
  {title} {\emph {Surface state dominated transport in topological insulator
  Bi2Te3 nanowires}},\ }\href {\doibase 10.1063/1.4829748} {\bibfield
  {journal} {\bibinfo  {journal} {Applied Physics Letters}\ }\textbf {\bibinfo
  {volume} {103}},\ \bibinfo {pages} {193107} (\bibinfo {year} {2013})},\
  \Eprint {http://arxiv.org/abs/https://doi.org/10.1063/1.4829748}
  {https://doi.org/10.1063/1.4829748} \BibitemShut {NoStop}%
\bibitem [{\citenamefont {Cho}\ \emph {et~al.}(2015)\citenamefont {Cho},
  \citenamefont {Dellabetta}, \citenamefont {Zhong}, \citenamefont
  {Schneeloch}, \citenamefont {Liu}, \citenamefont {Gu}, \citenamefont
  {Gilbert},\ and\ \citenamefont {Mason}}]{Rods2}%
  \BibitemOpen
  \bibfield  {author} {\bibinfo {author} {\bibfnamefont {S.}~\bibnamefont
  {Cho}}, \bibinfo {author} {\bibfnamefont {B.}~\bibnamefont {Dellabetta}},
  \bibinfo {author} {\bibfnamefont {R.}~\bibnamefont {Zhong}}, \bibinfo
  {author} {\bibfnamefont {J.}~\bibnamefont {Schneeloch}}, \bibinfo {author}
  {\bibfnamefont {T.}~\bibnamefont {Liu}}, \bibinfo {author} {\bibfnamefont
  {G.}~\bibnamefont {Gu}}, \bibinfo {author} {\bibfnamefont {M.~J.}\
  \bibnamefont {Gilbert}}, \ and\ \bibinfo {author} {\bibfnamefont
  {N.}~\bibnamefont {Mason}},\ }\bibfield  {title} {\bibinfo {title} {\emph
  {Aharonov--Bohm oscillations in a quasi-ballistic three-dimensional
  topological insulator nanowire}},\ }\href {\doibase 10.1038/ncomms8634}
  {\bibfield  {journal} {\bibinfo  {journal} {Nature Communications}\ }\textbf
  {\bibinfo {volume} {6}},\ \bibinfo {pages} {7634} (\bibinfo {year}
  {2015})}\BibitemShut {NoStop}%
\bibitem [{\citenamefont {M{\"u}nning}\ \emph {et~al.}(2021)\citenamefont
  {M{\"u}nning}, \citenamefont {Breunig}, \citenamefont {Legg}, \citenamefont
  {Roitsch}, \citenamefont {Fan}, \citenamefont {R{\"o}{\ss}ler}, \citenamefont
  {Rosch},\ and\ \citenamefont {Ando}}]{Rods3}%
  \BibitemOpen
  \bibfield  {author} {\bibinfo {author} {\bibfnamefont {F.}~\bibnamefont
  {M{\"u}nning}}, \bibinfo {author} {\bibfnamefont {O.}~\bibnamefont
  {Breunig}}, \bibinfo {author} {\bibfnamefont {H.~F.}\ \bibnamefont {Legg}},
  \bibinfo {author} {\bibfnamefont {S.}~\bibnamefont {Roitsch}}, \bibinfo
  {author} {\bibfnamefont {D.}~\bibnamefont {Fan}}, \bibinfo {author}
  {\bibfnamefont {M.}~\bibnamefont {R{\"o}{\ss}ler}}, \bibinfo {author}
  {\bibfnamefont {A.}~\bibnamefont {Rosch}}, \ and\ \bibinfo {author}
  {\bibfnamefont {Y.}~\bibnamefont {Ando}},\ }\bibfield  {title} {\bibinfo
  {title} {\emph {Quantum confinement of the Dirac surface states in
  topological-insulator nanowires}},\ }\href {\doibase
  10.1038/s41467-021-21230-3} {\bibfield  {journal} {\bibinfo  {journal}
  {Nature Communications}\ }\textbf {\bibinfo {volume} {12}},\ \bibinfo {pages}
  {1038} (\bibinfo {year} {2021})}\BibitemShut {NoStop}%
\bibitem [{\citenamefont {Kunakova}\ \emph {et~al.}(2018)\citenamefont
  {Kunakova}, \citenamefont {Galletti}, \citenamefont {Charpentier},
  \citenamefont {Andzane}, \citenamefont {Erts}, \citenamefont {Léonard},
  \citenamefont {Spataru}, \citenamefont {Bauch},\ and\ \citenamefont
  {Lombardi}}]{Rods4}%
  \BibitemOpen
  \bibfield  {author} {\bibinfo {author} {\bibfnamefont {G.}~\bibnamefont
  {Kunakova}}, \bibinfo {author} {\bibfnamefont {L.}~\bibnamefont {Galletti}},
  \bibinfo {author} {\bibfnamefont {S.}~\bibnamefont {Charpentier}}, \bibinfo
  {author} {\bibfnamefont {J.}~\bibnamefont {Andzane}}, \bibinfo {author}
  {\bibfnamefont {D.}~\bibnamefont {Erts}}, \bibinfo {author} {\bibfnamefont
  {F.}~\bibnamefont {Léonard}}, \bibinfo {author} {\bibfnamefont {C.~D.}\
  \bibnamefont {Spataru}}, \bibinfo {author} {\bibfnamefont {T.}~\bibnamefont
  {Bauch}}, \ and\ \bibinfo {author} {\bibfnamefont {F.}~\bibnamefont
  {Lombardi}},\ }\bibfield  {title} {\bibinfo {title} {\emph {Bulk-free
  topological insulator Bi2Se3 nanoribbons with magnetotransport signatures of
  Dirac surface states}},\ }\href {\doibase 10.1039/C8NR05500A} {\bibfield
  {journal} {\bibinfo  {journal} {Nanoscale}\ }\textbf {\bibinfo {volume}
  {10}},\ \bibinfo {pages} {19595} (\bibinfo {year} {2018})}\BibitemShut
  {NoStop}%
\bibitem [{\citenamefont {Ziegler}\ \emph {et~al.}(2018)\citenamefont
  {Ziegler}, \citenamefont {Kozlovsky}, \citenamefont {Gorini}, \citenamefont
  {Liu}, \citenamefont {Weish\"aupl}, \citenamefont {Maier}, \citenamefont
  {Fischer}, \citenamefont {Kozlov}, \citenamefont {Kvon}, \citenamefont
  {Mikhailov}, \citenamefont {Dvoretsky}, \citenamefont {Richter},\ and\
  \citenamefont {Weiss}}]{Rods5}%
  \BibitemOpen
  \bibfield  {author} {\bibinfo {author} {\bibfnamefont {J.}~\bibnamefont
  {Ziegler}}, \bibinfo {author} {\bibfnamefont {R.}~\bibnamefont {Kozlovsky}},
  \bibinfo {author} {\bibfnamefont {C.}~\bibnamefont {Gorini}}, \bibinfo
  {author} {\bibfnamefont {M.-H.}\ \bibnamefont {Liu}}, \bibinfo {author}
  {\bibfnamefont {S.}~\bibnamefont {Weish\"aupl}}, \bibinfo {author}
  {\bibfnamefont {H.}~\bibnamefont {Maier}}, \bibinfo {author} {\bibfnamefont
  {R.}~\bibnamefont {Fischer}}, \bibinfo {author} {\bibfnamefont {D.~A.}\
  \bibnamefont {Kozlov}}, \bibinfo {author} {\bibfnamefont {Z.~D.}\
  \bibnamefont {Kvon}}, \bibinfo {author} {\bibfnamefont {N.}~\bibnamefont
  {Mikhailov}}, \bibinfo {author} {\bibfnamefont {S.~A.}\ \bibnamefont
  {Dvoretsky}}, \bibinfo {author} {\bibfnamefont {K.}~\bibnamefont {Richter}},
  \ and\ \bibinfo {author} {\bibfnamefont {D.}~\bibnamefont {Weiss}},\
  }\bibfield  {title} {\bibinfo {title} {\emph {Probing spin helical surface
  states in topological HgTe nanowires}},\ }\href {\doibase
  10.1103/PhysRevB.97.035157} {\bibfield  {journal} {\bibinfo  {journal} {Phys.
  Rev. B}\ }\textbf {\bibinfo {volume} {97}},\ \bibinfo {pages} {035157}
  (\bibinfo {year} {2018})}\BibitemShut {NoStop}%
\bibitem [{\citenamefont {Jauregui}\ \emph {et~al.}(2015)\citenamefont
  {Jauregui}, \citenamefont {Pettes}, \citenamefont {Rokhinson}, \citenamefont
  {Shi},\ and\ \citenamefont {Chen}}]{Rods6}%
  \BibitemOpen
  \bibfield  {author} {\bibinfo {author} {\bibfnamefont {L.~A.}\ \bibnamefont
  {Jauregui}}, \bibinfo {author} {\bibfnamefont {M.~T.}\ \bibnamefont
  {Pettes}}, \bibinfo {author} {\bibfnamefont {L.~P.}\ \bibnamefont
  {Rokhinson}}, \bibinfo {author} {\bibfnamefont {L.}~\bibnamefont {Shi}}, \
  and\ \bibinfo {author} {\bibfnamefont {Y.~P.}\ \bibnamefont {Chen}},\
  }\bibfield  {title} {\bibinfo {title} {\emph {Gate Tunable Relativistic Mass
  and Berry's phase in Topological Insulator Nanoribbon Field Effect
  Devices}},\ }\href {\doibase 10.1038/srep08452} {\bibfield  {journal}
  {\bibinfo  {journal} {Scientific Reports}\ }\textbf {\bibinfo {volume} {5}},\
  \bibinfo {pages} {8452} (\bibinfo {year} {2015})}\BibitemShut {NoStop}%
\bibitem [{\citenamefont {Dufouleur}\ \emph {et~al.}(2017)\citenamefont
  {Dufouleur}, \citenamefont {Veyrat}, \citenamefont {Dassonneville},
  \citenamefont {Xypakis}, \citenamefont {Bardarson}, \citenamefont {Nowka},
  \citenamefont {Hampel}, \citenamefont {Schumann}, \citenamefont {Eichler},
  \citenamefont {Schmidt}, \citenamefont {B{\"u}chner},\ and\ \citenamefont
  {Giraud}}]{Rods7}%
  \BibitemOpen
  \bibfield  {author} {\bibinfo {author} {\bibfnamefont {J.}~\bibnamefont
  {Dufouleur}}, \bibinfo {author} {\bibfnamefont {L.}~\bibnamefont {Veyrat}},
  \bibinfo {author} {\bibfnamefont {B.}~\bibnamefont {Dassonneville}}, \bibinfo
  {author} {\bibfnamefont {E.}~\bibnamefont {Xypakis}}, \bibinfo {author}
  {\bibfnamefont {J.~H.}\ \bibnamefont {Bardarson}}, \bibinfo {author}
  {\bibfnamefont {C.}~\bibnamefont {Nowka}}, \bibinfo {author} {\bibfnamefont
  {S.}~\bibnamefont {Hampel}}, \bibinfo {author} {\bibfnamefont
  {J.}~\bibnamefont {Schumann}}, \bibinfo {author} {\bibfnamefont
  {B.}~\bibnamefont {Eichler}}, \bibinfo {author} {\bibfnamefont {O.~G.}\
  \bibnamefont {Schmidt}}, \bibinfo {author} {\bibfnamefont {B.}~\bibnamefont
  {B{\"u}chner}}, \ and\ \bibinfo {author} {\bibfnamefont {R.}~\bibnamefont
  {Giraud}},\ }\bibfield  {title} {\bibinfo {title} {\emph {Weakly-coupled
  quasi-1D helical modes in disordered 3D topological insulator quantum
  wires}},\ }\href {\doibase 10.1038/srep45276} {\bibfield  {journal} {\bibinfo
   {journal} {Scientific Reports}\ }\textbf {\bibinfo {volume} {7}},\ \bibinfo
  {pages} {45276} (\bibinfo {year} {2017})}\BibitemShut {NoStop}%
\bibitem [{\citenamefont {Autore}\ \emph {et~al.}(2015)\citenamefont {Autore},
  \citenamefont {Engelkamp}, \citenamefont {D'Apuzzo}, \citenamefont {Gaspare},
  \citenamefont {Pietro}, \citenamefont {Vecchio}, \citenamefont {Brahlek},
  \citenamefont {Koirala}, \citenamefont {Oh},\ and\ \citenamefont
  {Lupi}}]{PlasmonsTI}%
  \BibitemOpen
  \bibfield  {author} {\bibinfo {author} {\bibfnamefont {M.}~\bibnamefont
  {Autore}}, \bibinfo {author} {\bibfnamefont {H.}~\bibnamefont {Engelkamp}},
  \bibinfo {author} {\bibfnamefont {F.}~\bibnamefont {D'Apuzzo}}, \bibinfo
  {author} {\bibfnamefont {A.~D.}\ \bibnamefont {Gaspare}}, \bibinfo {author}
  {\bibfnamefont {P.~D.}\ \bibnamefont {Pietro}}, \bibinfo {author}
  {\bibfnamefont {I.~L.}\ \bibnamefont {Vecchio}}, \bibinfo {author}
  {\bibfnamefont {M.}~\bibnamefont {Brahlek}}, \bibinfo {author} {\bibfnamefont
  {N.}~\bibnamefont {Koirala}}, \bibinfo {author} {\bibfnamefont
  {S.}~\bibnamefont {Oh}}, \ and\ \bibinfo {author} {\bibfnamefont
  {S.}~\bibnamefont {Lupi}},\ }\bibfield  {title} {\bibinfo {title} {\emph
  {Observation of Magnetoplasmons in Bi2Se3 Topological Insulator}},\ }\href
  {\doibase 10.1021/acsphotonics.5b00036} {\bibfield  {journal} {\bibinfo
  {journal} {ACS Photonics}\ }\textbf {\bibinfo {volume} {2}},\ \bibinfo
  {pages} {1231} (\bibinfo {year} {2015})}\BibitemShut {NoStop}%
\bibitem [{Note11()}]{Note11}%
  \BibitemOpen
  \bibinfo {note} {In this regime, MP waves hybridize with light and become
  polaritons.}\BibitemShut {Stop}%
\bibitem [{\citenamefont {Bernstein}(1958)}]{BernsteinTheoryClassic1}%
  \BibitemOpen
  \bibfield  {author} {\bibinfo {author} {\bibfnamefont {I.~B.}\ \bibnamefont
  {Bernstein}},\ }\bibfield  {title} {\bibinfo {title} {\emph {Waves in a
  Plasma in a Magnetic Field}},\ }\href {\doibase 10.1103/PhysRev.109.10}
  {\bibfield  {journal} {\bibinfo  {journal} {Phys. Rev.}\ }\textbf {\bibinfo
  {volume} {109}},\ \bibinfo {pages} {10} (\bibinfo {year} {1958})}\BibitemShut
  {NoStop}%
\bibitem [{\citenamefont {Chiu}\ and\ \citenamefont
  {Quinn}(1974)}]{BernsteinTheoryClassic2}%
  \BibitemOpen
  \bibfield  {author} {\bibinfo {author} {\bibfnamefont {K.~W.}\ \bibnamefont
  {Chiu}}\ and\ \bibinfo {author} {\bibfnamefont {J.~J.}\ \bibnamefont
  {Quinn}},\ }\bibfield  {title} {\bibinfo {title} {\emph {Plasma oscillations
  of a two-dimensional electron gas in a strong magnetic field}},\ }\href
  {\doibase 10.1103/PhysRevB.9.4724} {\bibfield  {journal} {\bibinfo  {journal}
  {Phys. Rev. B}\ }\textbf {\bibinfo {volume} {9}},\ \bibinfo {pages} {4724}
  (\bibinfo {year} {1974})}\BibitemShut {NoStop}%
\bibitem [{\citenamefont {Volkov}\ and\ \citenamefont
  {Zabolotnykh}(2014)}]{BernsteinTheoryGraphene1}%
  \BibitemOpen
  \bibfield  {author} {\bibinfo {author} {\bibfnamefont {V.~A.}\ \bibnamefont
  {Volkov}}\ and\ \bibinfo {author} {\bibfnamefont {A.~A.}\ \bibnamefont
  {Zabolotnykh}},\ }\bibfield  {title} {\bibinfo {title} {\emph {Bernstein
  modes and giant microwave response of a two-dimensional electron system}},\
  }\href {\doibase 10.1103/PhysRevB.89.121410} {\bibfield  {journal} {\bibinfo
  {journal} {Phys. Rev. B}\ }\textbf {\bibinfo {volume} {89}},\ \bibinfo
  {pages} {121410} (\bibinfo {year} {2014})}\BibitemShut {NoStop}%
\bibitem [{\citenamefont {Rold\'an}\ \emph {et~al.}(2011)\citenamefont
  {Rold\'an}, \citenamefont {Goerbig},\ and\ \citenamefont
  {Fuchs}}]{BernsteinTheoryGraphene2}%
  \BibitemOpen
  \bibfield  {author} {\bibinfo {author} {\bibfnamefont {R.}~\bibnamefont
  {Rold\'an}}, \bibinfo {author} {\bibfnamefont {M.~O.}\ \bibnamefont
  {Goerbig}}, \ and\ \bibinfo {author} {\bibfnamefont {J.-N.}\ \bibnamefont
  {Fuchs}},\ }\bibfield  {title} {\bibinfo {title} {\emph {Theory of Bernstein
  modes in graphene}},\ }\href {\doibase 10.1103/PhysRevB.83.205406} {\bibfield
   {journal} {\bibinfo  {journal} {Phys. Rev. B}\ }\textbf {\bibinfo {volume}
  {83}},\ \bibinfo {pages} {205406} (\bibinfo {year} {2011})}\BibitemShut
  {NoStop}%
\bibitem [{\citenamefont {Batke}\ \emph {et~al.}(1985)\citenamefont {Batke},
  \citenamefont {Heitmann}, \citenamefont {Kotthaus},\ and\ \citenamefont
  {Ploog}}]{BersteinModesExp1}%
  \BibitemOpen
  \bibfield  {author} {\bibinfo {author} {\bibfnamefont {E.}~\bibnamefont
  {Batke}}, \bibinfo {author} {\bibfnamefont {D.}~\bibnamefont {Heitmann}},
  \bibinfo {author} {\bibfnamefont {J.~P.}\ \bibnamefont {Kotthaus}}, \ and\
  \bibinfo {author} {\bibfnamefont {K.}~\bibnamefont {Ploog}},\ }\bibfield
  {title} {\bibinfo {title} {\emph {Nonlocality in the Two-Dimensional Plasmon
  Dispersion}},\ }\href {\doibase 10.1103/PhysRevLett.54.2367} {\bibfield
  {journal} {\bibinfo  {journal} {Phys. Rev. Lett.}\ }\textbf {\bibinfo
  {volume} {54}},\ \bibinfo {pages} {2367} (\bibinfo {year}
  {1985})}\BibitemShut {NoStop}%
\bibitem [{\citenamefont {Batke}\ \emph {et~al.}(1986)\citenamefont {Batke},
  \citenamefont {Heitmann},\ and\ \citenamefont {Tu}}]{BersteinModesExp2}%
  \BibitemOpen
  \bibfield  {author} {\bibinfo {author} {\bibfnamefont {E.}~\bibnamefont
  {Batke}}, \bibinfo {author} {\bibfnamefont {D.}~\bibnamefont {Heitmann}}, \
  and\ \bibinfo {author} {\bibfnamefont {C.~W.}\ \bibnamefont {Tu}},\
  }\bibfield  {title} {\bibinfo {title} {\emph {Plasmon and magnetoplasmon
  excitation in two-dimensional electron space-charge layers on GaAs}},\ }\href
  {\doibase 10.1103/PhysRevB.34.6951} {\bibfield  {journal} {\bibinfo
  {journal} {Phys. Rev. B}\ }\textbf {\bibinfo {volume} {34}},\ \bibinfo
  {pages} {6951} (\bibinfo {year} {1986})}\BibitemShut {NoStop}%
\bibitem [{\citenamefont {Holland}\ \emph {et~al.}(2004)\citenamefont
  {Holland}, \citenamefont {Heyn}, \citenamefont {Heitmann}, \citenamefont
  {Batke}, \citenamefont {Hey}, \citenamefont {Friedland},\ and\ \citenamefont
  {Hu}}]{BersteinModesExp3}%
  \BibitemOpen
  \bibfield  {author} {\bibinfo {author} {\bibfnamefont {S.}~\bibnamefont
  {Holland}}, \bibinfo {author} {\bibfnamefont {C.}~\bibnamefont {Heyn}},
  \bibinfo {author} {\bibfnamefont {D.}~\bibnamefont {Heitmann}}, \bibinfo
  {author} {\bibfnamefont {E.}~\bibnamefont {Batke}}, \bibinfo {author}
  {\bibfnamefont {R.}~\bibnamefont {Hey}}, \bibinfo {author} {\bibfnamefont
  {K.~J.}\ \bibnamefont {Friedland}}, \ and\ \bibinfo {author} {\bibfnamefont
  {C.-M.}\ \bibnamefont {Hu}},\ }\bibfield  {title} {\bibinfo {title} {\emph
  {Quantized Dispersion of Two-Dimensional Magnetoplasmons Detected by
  Photoconductivity Spectroscopy}},\ }\href {\doibase
  10.1103/PhysRevLett.93.186804} {\bibfield  {journal} {\bibinfo  {journal}
  {Phys. Rev. Lett.}\ }\textbf {\bibinfo {volume} {93}},\ \bibinfo {pages}
  {186804} (\bibinfo {year} {2004})}\BibitemShut {NoStop}%
\bibitem [{\citenamefont {Richards}(2000)}]{BersteinModesExp4}%
  \BibitemOpen
  \bibfield  {author} {\bibinfo {author} {\bibfnamefont {D.}~\bibnamefont
  {Richards}},\ }\bibfield  {title} {\bibinfo {title} {\emph {Inelastic light
  scattering from inter-Landau level excitations in a two-dimensional electron
  gas}},\ }\href {\doibase 10.1103/PhysRevB.61.7517} {\bibfield  {journal}
  {\bibinfo  {journal} {Phys. Rev. B}\ }\textbf {\bibinfo {volume} {61}},\
  \bibinfo {pages} {7517} (\bibinfo {year} {2000})}\BibitemShut {NoStop}%
\bibitem [{\citenamefont {Bandurin}\ \emph {et~al.}(2022)\citenamefont
  {Bandurin}, \citenamefont {M{\"o}nch}, \citenamefont {Kapralov},
  \citenamefont {Phinney}, \citenamefont {Lindner}, \citenamefont {Liu},
  \citenamefont {Edgar}, \citenamefont {Dmitriev}, \citenamefont
  {Jarillo-Herrero}, \citenamefont {Svintsov},\ and\ \citenamefont
  {Ganichev}}]{BersteinModesExpGraphene}%
  \BibitemOpen
  \bibfield  {author} {\bibinfo {author} {\bibfnamefont {D.~A.}\ \bibnamefont
  {Bandurin}}, \bibinfo {author} {\bibfnamefont {E.}~\bibnamefont {M{\"o}nch}},
  \bibinfo {author} {\bibfnamefont {K.}~\bibnamefont {Kapralov}}, \bibinfo
  {author} {\bibfnamefont {I.~Y.}\ \bibnamefont {Phinney}}, \bibinfo {author}
  {\bibfnamefont {K.}~\bibnamefont {Lindner}}, \bibinfo {author} {\bibfnamefont
  {S.}~\bibnamefont {Liu}}, \bibinfo {author} {\bibfnamefont {J.~H.}\
  \bibnamefont {Edgar}}, \bibinfo {author} {\bibfnamefont {I.~A.}\ \bibnamefont
  {Dmitriev}}, \bibinfo {author} {\bibfnamefont {P.}~\bibnamefont
  {Jarillo-Herrero}}, \bibinfo {author} {\bibfnamefont {D.}~\bibnamefont
  {Svintsov}}, \ and\ \bibinfo {author} {\bibfnamefont {S.~D.}\ \bibnamefont
  {Ganichev}},\ }\bibfield  {title} {\bibinfo {title} {\emph {Cyclotron
  resonance overtones and near-field magnetoabsorption via terahertz Bernstein
  modes in graphene}},\ }\href {\doibase 10.1038/s41567-021-01494-8} {\bibfield
   {journal} {\bibinfo  {journal} {Nature Physics}\ }\textbf {\bibinfo {volume}
  {18}},\ \bibinfo {pages} {462} (\bibinfo {year} {2022})}\BibitemShut
  {NoStop}%
\bibitem [{\citenamefont {Allen}\ \emph {et~al.}(1983)\citenamefont {Allen},
  \citenamefont {St\"ormer},\ and\ \citenamefont {Hwang}}]{PlasmonsQW1}%
  \BibitemOpen
  \bibfield  {author} {\bibinfo {author} {\bibfnamefont {S.~J.}\ \bibnamefont
  {Allen}}, \bibinfo {author} {\bibfnamefont {H.~L.}\ \bibnamefont
  {St\"ormer}}, \ and\ \bibinfo {author} {\bibfnamefont {J.~C.~M.}\
  \bibnamefont {Hwang}},\ }\bibfield  {title} {\bibinfo {title} {\emph
  {Dimensional resonance of the two-dimensional electron gas in selectively
  doped GaAs/AlGaAs heterostructures}},\ }\href {\doibase
  10.1103/PhysRevB.28.4875} {\bibfield  {journal} {\bibinfo  {journal} {Phys.
  Rev. B}\ }\textbf {\bibinfo {volume} {28}},\ \bibinfo {pages} {4875}
  (\bibinfo {year} {1983})}\BibitemShut {NoStop}%
\bibitem [{\citenamefont {Kumada}\ \emph {et~al.}(2014)\citenamefont {Kumada},
  \citenamefont {Roulleau}, \citenamefont {Roche}, \citenamefont {Hashisaka},
  \citenamefont {Hibino}, \citenamefont {Petkovi\ifmmode~\acute{c}\else
  \'{c}\fi{}},\ and\ \citenamefont {Glattli}}]{PlasmonsGraphene1}%
  \BibitemOpen
  \bibfield  {author} {\bibinfo {author} {\bibfnamefont {N.}~\bibnamefont
  {Kumada}}, \bibinfo {author} {\bibfnamefont {P.}~\bibnamefont {Roulleau}},
  \bibinfo {author} {\bibfnamefont {B.}~\bibnamefont {Roche}}, \bibinfo
  {author} {\bibfnamefont {M.}~\bibnamefont {Hashisaka}}, \bibinfo {author}
  {\bibfnamefont {H.}~\bibnamefont {Hibino}}, \bibinfo {author} {\bibfnamefont
  {I.}~\bibnamefont {Petkovi\ifmmode~\acute{c}\else \'{c}\fi{}}}, \ and\
  \bibinfo {author} {\bibfnamefont {D.~C.}\ \bibnamefont {Glattli}},\
  }\bibfield  {title} {\bibinfo {title} {\emph {Resonant Edge Magnetoplasmons
  and Their Decay in Graphene}},\ }\href {\doibase
  10.1103/PhysRevLett.113.266601} {\bibfield  {journal} {\bibinfo  {journal}
  {Phys. Rev. Lett.}\ }\textbf {\bibinfo {volume} {113}},\ \bibinfo {pages}
  {266601} (\bibinfo {year} {2014})}\BibitemShut {NoStop}%
\bibitem [{\citenamefont {Crassee}\ \emph {et~al.}(2012)\citenamefont
  {Crassee}, \citenamefont {Orlita}, \citenamefont {Potemski}, \citenamefont
  {Walter}, \citenamefont {Ostler}, \citenamefont {Seyller}, \citenamefont
  {Gaponenko}, \citenamefont {Chen},\ and\ \citenamefont
  {Kuzmenko}}]{PlasmonsGraphene2}%
  \BibitemOpen
  \bibfield  {author} {\bibinfo {author} {\bibfnamefont {I.}~\bibnamefont
  {Crassee}}, \bibinfo {author} {\bibfnamefont {M.}~\bibnamefont {Orlita}},
  \bibinfo {author} {\bibfnamefont {M.}~\bibnamefont {Potemski}}, \bibinfo
  {author} {\bibfnamefont {A.~L.}\ \bibnamefont {Walter}}, \bibinfo {author}
  {\bibfnamefont {M.}~\bibnamefont {Ostler}}, \bibinfo {author} {\bibfnamefont
  {T.}~\bibnamefont {Seyller}}, \bibinfo {author} {\bibfnamefont
  {I.}~\bibnamefont {Gaponenko}}, \bibinfo {author} {\bibfnamefont
  {J.}~\bibnamefont {Chen}}, \ and\ \bibinfo {author} {\bibfnamefont {A.~B.}\
  \bibnamefont {Kuzmenko}},\ }\bibfield  {title} {\bibinfo {title} {\emph
  {Intrinsic Terahertz Plasmons and Magnetoplasmons in Large Scale Monolayer
  Graphene}},\ }\href {\doibase 10.1021/nl300572y} {\bibfield  {journal}
  {\bibinfo  {journal} {Nano Letters}\ }\textbf {\bibinfo {volume} {12}},\
  \bibinfo {pages} {2470} (\bibinfo {year} {2012})}\BibitemShut {NoStop}%
\bibitem [{\citenamefont {Yan}\ \emph {et~al.}(2012)\citenamefont {Yan},
  \citenamefont {Li}, \citenamefont {Li}, \citenamefont {Zhu}, \citenamefont
  {Avouris},\ and\ \citenamefont {Xia}}]{PlasmonsGraphene3}%
  \BibitemOpen
  \bibfield  {author} {\bibinfo {author} {\bibfnamefont {H.}~\bibnamefont
  {Yan}}, \bibinfo {author} {\bibfnamefont {Z.}~\bibnamefont {Li}}, \bibinfo
  {author} {\bibfnamefont {X.}~\bibnamefont {Li}}, \bibinfo {author}
  {\bibfnamefont {W.}~\bibnamefont {Zhu}}, \bibinfo {author} {\bibfnamefont
  {P.}~\bibnamefont {Avouris}}, \ and\ \bibinfo {author} {\bibfnamefont
  {F.}~\bibnamefont {Xia}},\ }\bibfield  {title} {\bibinfo {title} {\emph
  {Infrared Spectroscopy of Tunable Dirac Terahertz Magneto-Plasmons in
  Graphene}},\ }\href {\doibase 10.1021/nl3016335} {\bibfield  {journal}
  {\bibinfo  {journal} {Nano Letters}\ }\textbf {\bibinfo {volume} {12}},\
  \bibinfo {pages} {3766} (\bibinfo {year} {2012})}\BibitemShut {NoStop}%
\bibitem [{\citenamefont {Jin}\ \emph {et~al.}(2019)\citenamefont {Jin},
  \citenamefont {Xia}, \citenamefont {Christensen}, \citenamefont {Freeman},
  \citenamefont {Wang}, \citenamefont {Fong}, \citenamefont {Gardner},
  \citenamefont {Fallahi}, \citenamefont {Hu}, \citenamefont {Wang},
  \citenamefont {Engel}, \citenamefont {Xiao}, \citenamefont {Manfra},
  \citenamefont {Fang},\ and\ \citenamefont {Zhang}}]{MagnetoPlasmons2}%
  \BibitemOpen
  \bibfield  {author} {\bibinfo {author} {\bibfnamefont {D.}~\bibnamefont
  {Jin}}, \bibinfo {author} {\bibfnamefont {Y.}~\bibnamefont {Xia}}, \bibinfo
  {author} {\bibfnamefont {T.}~\bibnamefont {Christensen}}, \bibinfo {author}
  {\bibfnamefont {M.}~\bibnamefont {Freeman}}, \bibinfo {author} {\bibfnamefont
  {S.}~\bibnamefont {Wang}}, \bibinfo {author} {\bibfnamefont {K.~Y.}\
  \bibnamefont {Fong}}, \bibinfo {author} {\bibfnamefont {G.~C.}\ \bibnamefont
  {Gardner}}, \bibinfo {author} {\bibfnamefont {S.}~\bibnamefont {Fallahi}},
  \bibinfo {author} {\bibfnamefont {Q.}~\bibnamefont {Hu}}, \bibinfo {author}
  {\bibfnamefont {Y.}~\bibnamefont {Wang}}, \bibinfo {author} {\bibfnamefont
  {L.}~\bibnamefont {Engel}}, \bibinfo {author} {\bibfnamefont {Z.-L.}\
  \bibnamefont {Xiao}}, \bibinfo {author} {\bibfnamefont {M.~J.}\ \bibnamefont
  {Manfra}}, \bibinfo {author} {\bibfnamefont {N.~X.}\ \bibnamefont {Fang}}, \
  and\ \bibinfo {author} {\bibfnamefont {X.}~\bibnamefont {Zhang}},\ }\bibfield
   {title} {\bibinfo {title} {\emph {Topological kink plasmons on
  magnetic-domain boundaries}},\ }\href {\doibase 10.1038/s41467-019-12092-x}
  {\bibfield  {journal} {\bibinfo  {journal} {Nature Communications}\ }\textbf
  {\bibinfo {volume} {10}},\ \bibinfo {pages} {4565} (\bibinfo {year}
  {2019})}\BibitemShut {NoStop}%
\bibitem [{\citenamefont {Tan}\ \emph {et~al.}(2008)\citenamefont {Tan},
  \citenamefont {Yuan}, \citenamefont {Lin}, \citenamefont {Wang},
  \citenamefont {Mei}, \citenamefont {Burge},\ and\ \citenamefont
  {Mu}}]{OpticalVorticities1}%
  \BibitemOpen
  \bibfield  {author} {\bibinfo {author} {\bibfnamefont {P.~S.}\ \bibnamefont
  {Tan}}, \bibinfo {author} {\bibfnamefont {X.-C.}\ \bibnamefont {Yuan}},
  \bibinfo {author} {\bibfnamefont {J.}~\bibnamefont {Lin}}, \bibinfo {author}
  {\bibfnamefont {Q.}~\bibnamefont {Wang}}, \bibinfo {author} {\bibfnamefont
  {T.}~\bibnamefont {Mei}}, \bibinfo {author} {\bibfnamefont {R.~E.}\
  \bibnamefont {Burge}}, \ and\ \bibinfo {author} {\bibfnamefont {G.~G.}\
  \bibnamefont {Mu}},\ }\bibfield  {title} {\bibinfo {title} {\emph {Surface
  plasmon polaritons generated by optical vortex beams}},\ }\href {\doibase
  10.1063/1.2890058} {\bibfield  {journal} {\bibinfo  {journal} {Applied
  Physics Letters}\ }\textbf {\bibinfo {volume} {92}},\ \bibinfo {pages}
  {111108} (\bibinfo {year} {2008})}\BibitemShut {NoStop}%
\bibitem [{\citenamefont {Sakai}\ \emph {et~al.}(2015)\citenamefont {Sakai},
  \citenamefont {Nomura}, \citenamefont {Yamamoto},\ and\ \citenamefont
  {Sasaki}}]{OpticalVorticities2}%
  \BibitemOpen
  \bibfield  {author} {\bibinfo {author} {\bibfnamefont {K.}~\bibnamefont
  {Sakai}}, \bibinfo {author} {\bibfnamefont {K.}~\bibnamefont {Nomura}},
  \bibinfo {author} {\bibfnamefont {T.}~\bibnamefont {Yamamoto}}, \ and\
  \bibinfo {author} {\bibfnamefont {K.}~\bibnamefont {Sasaki}},\ }\bibfield
  {title} {\bibinfo {title} {\emph {Excitation of Multipole Plasmons by Optical
  Vortex Beams}},\ }\href {\doibase 10.1038/srep08431} {\bibfield  {journal}
  {\bibinfo  {journal} {Scientific Reports}\ }\textbf {\bibinfo {volume} {5}},\
  \bibinfo {pages} {8431} (\bibinfo {year} {2015})}\BibitemShut {NoStop}%
\bibitem [{\citenamefont {Ge}\ \emph {et~al.}(2015)\citenamefont {Ge},
  \citenamefont {Han},\ and\ \citenamefont {Zi}}]{MieSphere1}%
  \BibitemOpen
  \bibfield  {author} {\bibinfo {author} {\bibfnamefont {L.}~\bibnamefont
  {Ge}}, \bibinfo {author} {\bibfnamefont {D.}~\bibnamefont {Han}}, \ and\
  \bibinfo {author} {\bibfnamefont {J.}~\bibnamefont {Zi}},\ }\bibfield
  {title} {\bibinfo {title} {\emph {Electromagnetic scattering by spheres of
  topological insulators}},\ }\href {\doibase
  https://doi.org/10.1016/j.optcom.2015.05.054} {\bibfield  {journal} {\bibinfo
   {journal} {Optics Communications}\ }\textbf {\bibinfo {volume} {354}},\
  \bibinfo {pages} {225} (\bibinfo {year} {2015})}\BibitemShut {NoStop}%
\bibitem [{\citenamefont {Lakhtakia}\ and\ \citenamefont
  {Mackay}(2016)}]{MieSphere2}%
  \BibitemOpen
  \bibfield  {author} {\bibinfo {author} {\bibfnamefont {A.}~\bibnamefont
  {Lakhtakia}}\ and\ \bibinfo {author} {\bibfnamefont {T.~G.}\ \bibnamefont
  {Mackay}},\ }\bibfield  {title} {\bibinfo {title} {\emph {Electromagnetic
  scattering by homogeneous, isotropic, dielectric--magnetic sphere with
  topologically insulating surface states}},\ }\href {\doibase
  10.1364/JOSAB.33.000603} {\bibfield  {journal} {\bibinfo  {journal} {J. Opt.
  Soc. Am. B}\ }\textbf {\bibinfo {volume} {33}},\ \bibinfo {pages} {603}
  (\bibinfo {year} {2016})}\BibitemShut {NoStop}%
\bibitem [{\citenamefont {Schultz}\ \emph {et~al.}(2020)\citenamefont
  {Schultz}, \citenamefont {Nogueira}, \citenamefont {Büchner}, \citenamefont
  {Brink},\ and\ \citenamefont {Lubk}}]{MieSphere3}%
  \BibitemOpen
  \bibfield  {author} {\bibinfo {author} {\bibfnamefont {J.}~\bibnamefont
  {Schultz}}, \bibinfo {author} {\bibfnamefont {F.~S.}\ \bibnamefont
  {Nogueira}}, \bibinfo {author} {\bibfnamefont {B.}~\bibnamefont {Büchner}},
  \bibinfo {author} {\bibfnamefont {J.~v.~d.}\ \bibnamefont {Brink}}, \ and\
  \bibinfo {author} {\bibfnamefont {A.}~\bibnamefont {Lubk}},\ }\href {\doibase
  10.48550/ARXIV.2002.03804} {\bibinfo {title} {\emph {Axion Mie Theory of
  Electron Energy Loss Spectroscopy in Topological Insulators}}} (\bibinfo
  {year} {2020})\BibitemShut {NoStop}%
\bibitem [{\citenamefont {Imura}\ \emph {et~al.}(2011)\citenamefont {Imura},
  \citenamefont {Takane},\ and\ \citenamefont {Tanaka}}]{ABCylinder1}%
  \BibitemOpen
  \bibfield  {author} {\bibinfo {author} {\bibfnamefont {K.-I.}\ \bibnamefont
  {Imura}}, \bibinfo {author} {\bibfnamefont {Y.}~\bibnamefont {Takane}}, \
  and\ \bibinfo {author} {\bibfnamefont {A.}~\bibnamefont {Tanaka}},\
  }\bibfield  {title} {\bibinfo {title} {\emph {Spin Berry phase in anisotropic
  topological insulators}},\ }\href {\doibase 10.1103/PhysRevB.84.195406}
  {\bibfield  {journal} {\bibinfo  {journal} {Phys. Rev. B}\ }\textbf {\bibinfo
  {volume} {84}},\ \bibinfo {pages} {195406} (\bibinfo {year}
  {2011})}\BibitemShut {NoStop}%
\bibitem [{\citenamefont {Governale}\ \emph {et~al.}(2020)\citenamefont
  {Governale}, \citenamefont {Bhandari}, \citenamefont {Taddei}, \citenamefont
  {Imura},\ and\ \citenamefont {Zülicke}}]{ABCylinder2}%
  \BibitemOpen
  \bibfield  {author} {\bibinfo {author} {\bibfnamefont {M.}~\bibnamefont
  {Governale}}, \bibinfo {author} {\bibfnamefont {B.}~\bibnamefont {Bhandari}},
  \bibinfo {author} {\bibfnamefont {F.}~\bibnamefont {Taddei}}, \bibinfo
  {author} {\bibfnamefont {K.-I.}\ \bibnamefont {Imura}}, \ and\ \bibinfo
  {author} {\bibfnamefont {U.}~\bibnamefont {Zülicke}},\ }\bibfield  {title}
  {\bibinfo {title} {\emph {Finite-size effects in cylindrical topological
  insulators}},\ }\href {\doibase 10.1088/1367-2630/ab90d3} {\bibfield
  {journal} {\bibinfo  {journal} {New Journal of Physics}\ }\textbf {\bibinfo
  {volume} {22}},\ \bibinfo {pages} {063042} (\bibinfo {year}
  {2020})}\BibitemShut {NoStop}%
\bibitem [{\citenamefont {Imura}\ \emph {et~al.}(2012)\citenamefont {Imura},
  \citenamefont {Yoshimura}, \citenamefont {Takane},\ and\ \citenamefont
  {Fukui}}]{ABSpherical1}%
  \BibitemOpen
  \bibfield  {author} {\bibinfo {author} {\bibfnamefont {K.-I.}\ \bibnamefont
  {Imura}}, \bibinfo {author} {\bibfnamefont {Y.}~\bibnamefont {Yoshimura}},
  \bibinfo {author} {\bibfnamefont {Y.}~\bibnamefont {Takane}}, \ and\ \bibinfo
  {author} {\bibfnamefont {T.}~\bibnamefont {Fukui}},\ }\bibfield  {title}
  {\bibinfo {title} {\emph {Spherical topological insulator}},\ }\href
  {\doibase 10.1103/PhysRevB.86.235119} {\bibfield  {journal} {\bibinfo
  {journal} {Phys. Rev. B}\ }\textbf {\bibinfo {volume} {86}},\ \bibinfo
  {pages} {235119} (\bibinfo {year} {2012})}\BibitemShut {NoStop}%
\bibitem [{\citenamefont {Gioia}\ \emph {et~al.}(2019)\citenamefont {Gioia},
  \citenamefont {Christie}, \citenamefont {Z\"ulicke}, \citenamefont
  {Governale},\ and\ \citenamefont {Sneyd}}]{ABSpherical2}%
  \BibitemOpen
  \bibfield  {author} {\bibinfo {author} {\bibfnamefont {L.}~\bibnamefont
  {Gioia}}, \bibinfo {author} {\bibfnamefont {M.~G.}\ \bibnamefont {Christie}},
  \bibinfo {author} {\bibfnamefont {U.}~\bibnamefont {Z\"ulicke}}, \bibinfo
  {author} {\bibfnamefont {M.}~\bibnamefont {Governale}}, \ and\ \bibinfo
  {author} {\bibfnamefont {A.~J.}\ \bibnamefont {Sneyd}},\ }\bibfield  {title}
  {\bibinfo {title} {\emph {Spherical topological insulator nanoparticles:
  Quantum size effects and optical transitions}},\ }\href {\doibase
  10.1103/PhysRevB.100.205417} {\bibfield  {journal} {\bibinfo  {journal}
  {Phys. Rev. B}\ }\textbf {\bibinfo {volume} {100}},\ \bibinfo {pages}
  {205417} (\bibinfo {year} {2019})}\BibitemShut {NoStop}%
\bibitem [{\citenamefont {Siroki}\ \emph {et~al.}(2016)\citenamefont {Siroki},
  \citenamefont {Lee}, \citenamefont {Haynes},\ and\ \citenamefont
  {Giannini}}]{ABSpherical3}%
  \BibitemOpen
  \bibfield  {author} {\bibinfo {author} {\bibfnamefont {G.}~\bibnamefont
  {Siroki}}, \bibinfo {author} {\bibfnamefont {D.~K.~K.}\ \bibnamefont {Lee}},
  \bibinfo {author} {\bibfnamefont {P.~D.}\ \bibnamefont {Haynes}}, \ and\
  \bibinfo {author} {\bibfnamefont {V.}~\bibnamefont {Giannini}},\ }\bibfield
  {title} {\bibinfo {title} {\emph {Single-electron induced surface plasmons on
  a topological nanoparticle}},\ }\href {\doibase 10.1038/ncomms12375}
  {\bibfield  {journal} {\bibinfo  {journal} {Nature Communications}\ }\textbf
  {\bibinfo {volume} {7}},\ \bibinfo {pages} {12375} (\bibinfo {year}
  {2016})}\BibitemShut {NoStop}%
\bibitem [{\citenamefont {Raghu}\ \emph {et~al.}(2010)\citenamefont {Raghu},
  \citenamefont {Chung}, \citenamefont {Qi},\ and\ \citenamefont
  {Zhang}}]{SpinPlasmonsRaghu}%
  \BibitemOpen
  \bibfield  {author} {\bibinfo {author} {\bibfnamefont {S.}~\bibnamefont
  {Raghu}}, \bibinfo {author} {\bibfnamefont {S.~B.}\ \bibnamefont {Chung}},
  \bibinfo {author} {\bibfnamefont {X.-L.}\ \bibnamefont {Qi}}, \ and\ \bibinfo
  {author} {\bibfnamefont {S.-C.}\ \bibnamefont {Zhang}},\ }\bibfield  {title}
  {\bibinfo {title} {\emph {Collective Modes of a Helical Liquid}},\ }\href
  {\doibase 10.1103/PhysRevLett.104.116401} {\bibfield  {journal} {\bibinfo
  {journal} {Phys. Rev. Lett.}\ }\textbf {\bibinfo {volume} {104}},\ \bibinfo
  {pages} {116401} (\bibinfo {year} {2010})}\BibitemShut {NoStop}%
\bibitem [{\citenamefont {Efimkin}\ \emph
  {et~al.}(2012{\natexlab{a}})\citenamefont {Efimkin}, \citenamefont
  {Lozovik},\ and\ \citenamefont {Sokolik}}]{SpinPlasmonsEfimkin1}%
  \BibitemOpen
  \bibfield  {author} {\bibinfo {author} {\bibfnamefont {D.~K.}\ \bibnamefont
  {Efimkin}}, \bibinfo {author} {\bibfnamefont {Y.~E.}\ \bibnamefont
  {Lozovik}}, \ and\ \bibinfo {author} {\bibfnamefont {A.~A.}\ \bibnamefont
  {Sokolik}},\ }\bibfield  {title} {\bibinfo {title} {\emph {Collective
  excitations on a surface of topological insulator}},\ }\href {\doibase
  10.1186/1556-276X-7-163} {\bibfield  {journal} {\bibinfo  {journal}
  {Nanoscale Research Letters}\ }\textbf {\bibinfo {volume} {7}},\ \bibinfo
  {pages} {163} (\bibinfo {year} {2012}{\natexlab{a}})}\BibitemShut {NoStop}%
\bibitem [{\citenamefont {Efimkin}\ \emph
  {et~al.}(2012{\natexlab{b}})\citenamefont {Efimkin}, \citenamefont
  {Lozovik},\ and\ \citenamefont {Sokolik}}]{SpinPlasmonsEfimkin2}%
  \BibitemOpen
  \bibfield  {author} {\bibinfo {author} {\bibfnamefont {D.}~\bibnamefont
  {Efimkin}}, \bibinfo {author} {\bibfnamefont {Y.}~\bibnamefont {Lozovik}}, \
  and\ \bibinfo {author} {\bibfnamefont {A.}~\bibnamefont {Sokolik}},\
  }\bibfield  {title} {\bibinfo {title} {\emph {Spin-plasmons in topological
  insulator}},\ }\href {\doibase https://doi.org/10.1016/j.jmmm.2012.02.102}
  {\bibfield  {journal} {\bibinfo  {journal} {Journal of Magnetism and Magnetic
  Materials}\ }\textbf {\bibinfo {volume} {324}},\ \bibinfo {pages} {3610}
  (\bibinfo {year} {2012}{\natexlab{b}})}\BibitemShut {NoStop}%
\bibitem [{\citenamefont {Stauber}\ \emph {et~al.}(2013)\citenamefont
  {Stauber}, \citenamefont {G\'omez-Santos},\ and\ \citenamefont
  {Brey}}]{SpinPlasmonsStauber1}%
  \BibitemOpen
  \bibfield  {author} {\bibinfo {author} {\bibfnamefont {T.}~\bibnamefont
  {Stauber}}, \bibinfo {author} {\bibfnamefont {G.}~\bibnamefont
  {G\'omez-Santos}}, \ and\ \bibinfo {author} {\bibfnamefont {L.}~\bibnamefont
  {Brey}},\ }\bibfield  {title} {\bibinfo {title} {\emph {Spin-charge
  separation of plasmonic excitations in thin topological insulators}},\ }\href
  {\doibase 10.1103/PhysRevB.88.205427} {\bibfield  {journal} {\bibinfo
  {journal} {Phys. Rev. B}\ }\textbf {\bibinfo {volume} {88}},\ \bibinfo
  {pages} {205427} (\bibinfo {year} {2013})}\BibitemShut {NoStop}%
\bibitem [{\citenamefont {Efimkin}\ and\ \citenamefont
  {Kargarian}(2021)}]{SpinPlasmonsEfimkin3}%
  \BibitemOpen
  \bibfield  {author} {\bibinfo {author} {\bibfnamefont {D.~K.}\ \bibnamefont
  {Efimkin}}\ and\ \bibinfo {author} {\bibfnamefont {M.}~\bibnamefont
  {Kargarian}},\ }\bibfield  {title} {\bibinfo {title} {\emph {Topological
  spin-plasma waves}},\ }\href {\doibase 10.1103/PhysRevB.104.075413}
  {\bibfield  {journal} {\bibinfo  {journal} {Phys. Rev. B}\ }\textbf {\bibinfo
  {volume} {104}},\ \bibinfo {pages} {075413} (\bibinfo {year}
  {2021})}\BibitemShut {NoStop}%
\bibitem [{\citenamefont {Lai}\ \emph {et~al.}(2014)\citenamefont {Lai},
  \citenamefont {Lin}, \citenamefont {Wu},\ and\ \citenamefont
  {Liu}}]{SpinPlasmonsReview1}%
  \BibitemOpen
  \bibfield  {author} {\bibinfo {author} {\bibfnamefont {Y.-P.}\ \bibnamefont
  {Lai}}, \bibinfo {author} {\bibfnamefont {I.-T.}\ \bibnamefont {Lin}},
  \bibinfo {author} {\bibfnamefont {K.-H.}\ \bibnamefont {Wu}}, \ and\ \bibinfo
  {author} {\bibfnamefont {J.-M.}\ \bibnamefont {Liu}},\ }\bibfield  {title}
  {\bibinfo {title} {\emph {Plasmonics in Topological Insulators}},\ }\href
  {\doibase 10.5772/58558} {\bibfield  {journal} {\bibinfo  {journal}
  {Nanomaterials and Nanotechnology}\ }\textbf {\bibinfo {volume} {4}},\
  \bibinfo {pages} {13} (\bibinfo {year} {2014})}\BibitemShut {NoStop}%
\bibitem [{\citenamefont {Averitt}\ \emph {et~al.}(1997)\citenamefont
  {Averitt}, \citenamefont {Sarkar},\ and\ \citenamefont
  {Halas}}]{MetalicNanoShell1}%
  \BibitemOpen
  \bibfield  {author} {\bibinfo {author} {\bibfnamefont {R.~D.}\ \bibnamefont
  {Averitt}}, \bibinfo {author} {\bibfnamefont {D.}~\bibnamefont {Sarkar}}, \
  and\ \bibinfo {author} {\bibfnamefont {N.~J.}\ \bibnamefont {Halas}},\
  }\bibfield  {title} {\bibinfo {title} {\emph {Plasmon Resonance Shifts of
  Au-Coated ${\mathrm{Au}}_{2}S$ Nanoshells: Insight into Multicomponent
  Nanoparticle Growth}},\ }\href {\doibase 10.1103/PhysRevLett.78.4217}
  {\bibfield  {journal} {\bibinfo  {journal} {Phys. Rev. Lett.}\ }\textbf
  {\bibinfo {volume} {78}},\ \bibinfo {pages} {4217} (\bibinfo {year}
  {1997})}\BibitemShut {NoStop}%
\bibitem [{\citenamefont {Daneshfar}\ and\ \citenamefont
  {Bazyari}(2014)}]{MetalicNanoShell2}%
  \BibitemOpen
  \bibfield  {author} {\bibinfo {author} {\bibfnamefont {N.}~\bibnamefont
  {Daneshfar}}\ and\ \bibinfo {author} {\bibfnamefont {K.}~\bibnamefont
  {Bazyari}},\ }\bibfield  {title} {\bibinfo {title} {\emph {Optical and
  spectral tunability of multilayer spherical and cylindrical nanoshells}},\
  }\href {\doibase 10.1007/s00339-013-8188-z} {\bibfield  {journal} {\bibinfo
  {journal} {Applied Physics A}\ }\textbf {\bibinfo {volume} {116}},\ \bibinfo
  {pages} {611} (\bibinfo {year} {2014})}\BibitemShut {NoStop}%
\bibitem [{\citenamefont {Perera}\ \emph {et~al.}(2020)\citenamefont {Perera},
  \citenamefont {Gunapala}, \citenamefont {Stockman},\ and\ \citenamefont
  {Premaratne}}]{MetalicNanoShell3}%
  \BibitemOpen
  \bibfield  {author} {\bibinfo {author} {\bibfnamefont {T.}~\bibnamefont
  {Perera}}, \bibinfo {author} {\bibfnamefont {S.~D.}\ \bibnamefont
  {Gunapala}}, \bibinfo {author} {\bibfnamefont {M.~I.}\ \bibnamefont
  {Stockman}}, \ and\ \bibinfo {author} {\bibfnamefont {M.}~\bibnamefont
  {Premaratne}},\ }\bibfield  {title} {\bibinfo {title} {\emph {Plasmonic
  Properties of Metallic Nanoshells in the Quantum Limit: From Single Particle
  Excitations to Plasmons}},\ }\href {\doibase 10.1021/acs.jpcc.0c10507}
  {\bibfield  {journal} {\bibinfo  {journal} {The Journal of Physical Chemistry
  C}\ }\textbf {\bibinfo {volume} {124}},\ \bibinfo {pages} {27694} (\bibinfo
  {year} {2020})}\BibitemShut {NoStop}%
\bibitem [{\citenamefont {Mikhailov}\ and\ \citenamefont
  {Volkov}(1992)}]{EdgeDensity1}%
  \BibitemOpen
  \bibfield  {author} {\bibinfo {author} {\bibfnamefont {S.~A.}\ \bibnamefont
  {Mikhailov}}\ and\ \bibinfo {author} {\bibfnamefont {V.~A.}\ \bibnamefont
  {Volkov}},\ }\bibfield  {title} {\bibinfo {title} {\emph {Inter-edge
  magnetoplasmons in inhomogeneous two-dimensional electron systems}},\ }\href
  {\doibase 10.1088/0953-8984/4/31/005} {\bibfield  {journal} {\bibinfo
  {journal} {Journal of Physics: Condensed Matter}\ }\textbf {\bibinfo {volume}
  {4}},\ \bibinfo {pages} {6523} (\bibinfo {year} {1992})}\BibitemShut
  {NoStop}%
\bibitem [{\citenamefont {Aleiner}\ and\ \citenamefont
  {Glazman}(1994)}]{EdgeDensity2}%
  \BibitemOpen
  \bibfield  {author} {\bibinfo {author} {\bibfnamefont {I.~L.}\ \bibnamefont
  {Aleiner}}\ and\ \bibinfo {author} {\bibfnamefont {L.~I.}\ \bibnamefont
  {Glazman}},\ }\bibfield  {title} {\bibinfo {title} {\emph {Novel edge
  excitations of two-dimensional electron liquid in a magnetic field}},\ }\href
  {\doibase 10.1103/PhysRevLett.72.2935} {\bibfield  {journal} {\bibinfo
  {journal} {Phys. Rev. Lett.}\ }\textbf {\bibinfo {volume} {72}},\ \bibinfo
  {pages} {2935} (\bibinfo {year} {1994})}\BibitemShut {NoStop}%
\bibitem [{\citenamefont {Xia}\ and\ \citenamefont
  {Quinn}(1994)}]{EdgeDensity3}%
  \BibitemOpen
  \bibfield  {author} {\bibinfo {author} {\bibfnamefont {X.}~\bibnamefont
  {Xia}}\ and\ \bibinfo {author} {\bibfnamefont {J.~J.}\ \bibnamefont
  {Quinn}},\ }\bibfield  {title} {\bibinfo {title} {\emph {Edge magnetoplasmons
  of two-dimensional electron-gas systems}},\ }\href {\doibase
  10.1103/PhysRevB.50.11187} {\bibfield  {journal} {\bibinfo  {journal} {Phys.
  Rev. B}\ }\textbf {\bibinfo {volume} {50}},\ \bibinfo {pages} {11187}
  (\bibinfo {year} {1994})}\BibitemShut {NoStop}%
\bibitem [{\citenamefont {Song}\ and\ \citenamefont
  {Rudner}(2016)}]{EdgeAnomalous1}%
  \BibitemOpen
  \bibfield  {author} {\bibinfo {author} {\bibfnamefont {J.~C.~W.}\
  \bibnamefont {Song}}\ and\ \bibinfo {author} {\bibfnamefont {M.~S.}\
  \bibnamefont {Rudner}},\ }\bibfield  {title} {\bibinfo {title} {\emph {Chiral
  plasmons without magnetic field}},\ }\href {\doibase 10.1073/pnas.1519086113}
  {\bibfield  {journal} {\bibinfo  {journal} {Proceedings of the National
  Academy of Sciences}\ }\textbf {\bibinfo {volume} {113}},\ \bibinfo {pages}
  {4658} (\bibinfo {year} {2016})},\ \Eprint
  {http://arxiv.org/abs/https://www.pnas.org/doi/pdf/10.1073/pnas.1519086113}
  {https://www.pnas.org/doi/pdf/10.1073/pnas.1519086113} \BibitemShut {NoStop}%
\bibitem [{\citenamefont {Petrov}(2021)}]{EdgeAnomalous2}%
  \BibitemOpen
  \bibfield  {author} {\bibinfo {author} {\bibfnamefont {A.~S.}\ \bibnamefont
  {Petrov}},\ }\bibfield  {title} {\bibinfo {title} {\emph {Plasmonic
  excitation for a tunable transmitter without magnetic field immune to
  backscattering}},\ }\href {\doibase 10.1103/PhysRevB.104.L241407} {\bibfield
  {journal} {\bibinfo  {journal} {Phys. Rev. B}\ }\textbf {\bibinfo {volume}
  {104}},\ \bibinfo {pages} {L241407} (\bibinfo {year} {2021})}\BibitemShut
  {NoStop}%
\bibitem [{\citenamefont {Sokolik}\ \emph {et~al.}(2021)\citenamefont
  {Sokolik}, \citenamefont {Kotov},\ and\ \citenamefont
  {Lozovik}}]{Edge2DAnisotropic}%
  \BibitemOpen
  \bibfield  {author} {\bibinfo {author} {\bibfnamefont {A.~A.}\ \bibnamefont
  {Sokolik}}, \bibinfo {author} {\bibfnamefont {O.~V.}\ \bibnamefont {Kotov}},
  \ and\ \bibinfo {author} {\bibfnamefont {Y.~E.}\ \bibnamefont {Lozovik}},\
  }\bibfield  {title} {\bibinfo {title} {\emph {Plasmonic modes at inclined
  edges of anisotropic two-dimensional materials}},\ }\href {\doibase
  10.1103/PhysRevB.103.155402} {\bibfield  {journal} {\bibinfo  {journal}
  {Phys. Rev. B}\ }\textbf {\bibinfo {volume} {103}},\ \bibinfo {pages}
  {155402} (\bibinfo {year} {2021})}\BibitemShut {NoStop}%
\bibitem [{\citenamefont {Jackiw}\ and\ \citenamefont
  {Rebbi}(1976)}]{JackiwRebbi}%
  \BibitemOpen
  \bibfield  {author} {\bibinfo {author} {\bibfnamefont {R.}~\bibnamefont
  {Jackiw}}\ and\ \bibinfo {author} {\bibfnamefont {C.}~\bibnamefont {Rebbi}},\
  }\bibfield  {title} {\bibinfo {title} {\emph {Solitons with fermion number
  \textonehalf{}}},\ }\href {\doibase 10.1103/PhysRevD.13.3398} {\bibfield
  {journal} {\bibinfo  {journal} {Phys. Rev. D}\ }\textbf {\bibinfo {volume}
  {13}},\ \bibinfo {pages} {3398} (\bibinfo {year} {1976})}\BibitemShut
  {NoStop}%
\end{thebibliography}%
\bibliographystyle{apsrev4-1_our_style}

\newpage
\begin{widetext}

\begin{appendix}

\vspace{-0.15in}
\section{APPENDIX A. Robustness of equatorial MP modes in the cylindrical geometry}
\vspace{-0.05in}
\subsection{A1. Wave functions for Kelvin and Yanai modes}
Wave functions for Kelvin and Yanai MP modes can be directly calculated from the Hermitian eigenvalue problem defined in  Eq.~(\ref{HamiltonianFullPrime}). If we use the following ansatz: 
\begin{equation}
\label{WaveFunctionAnzatz}
\vec{j}^{x(y)}({\vec{r},t})=J^{x(y)}(x) e^{i (k y-\omega t)}, \quad \quad \quad \vec{j}^{0}({\vec{r},t})=J^{0}(x) e^{i (k y-\omega t)},
\end{equation}
the eigenvalue problem can be presented as 
\begin{align}
\label{WFEq1}
-i \omega J^x &=- u\partial_x J^0+ \omega_0 b(x) J^y, \\ \label{WFEq2} -i \omega J^0 &=- u\partial_x J^x - i u k  J^y, \\ \label{WFEq3} -i \omega J^y &=- i u k J^0 - \omega_0 b(x) J^x.
\end{align}
The Kelvin mode is longitudinal $J^x=0$, the transverse current profile is given by Eq.~(\ref{KelvinMode}), and the resulting density oscillation profile is equal to $J^{0}=-J^y$. If we incorporate the explicit expression for the domain wall profile $b(x)=\tanh(x/d)$, the wave function for the Kelvin mode (up to a normalization factor) can be presented as
\begin{equation}
\label{WFKelvin}
\psi_\mathrm{K}\propto\begin{pmatrix}
-\frac{i}{\sqrt{2}} \\
1  \\
\frac{i}{\sqrt{2}}  \\
\end{pmatrix} \frac{1}{\left[\cosh\left(\frac{x}{d}\right)\right]^{\frac{1}{\alpha}}},  
\end{equation}
where $\alpha=\omega_\mathrm{d}/\omega_0$ is the only controlling parameter introduced in the main text. Importantly, the profiles for its nonzero components $J^y$ and $J^0$ are symmetric across the domain wall. 

For the Yanai mode, the transverse current profile is given by $J^x\propto1/[\cosh\left(x/d\right)]^\lambda$. Here, $\lambda$ is wave-vector-dependent and must be evaluated from Eq.~(\ref{Eigenvalue3}). The longitudinal current and density profiles can be calculated from Eqs.~(\ref{WFEq2}) and (\ref{WFEq3}), which results in 
\begin{align}
\label{WFEq1}
J^y=\frac{i(u^2k\partial_x+\omega \omega_0 b(x))J^x}{u^2 k^2-\omega^2}=\frac{i \omega_0(\omega-u k\lambda \alpha)}{u^2 k^2-\omega^2} b(x) J^x = i A^y b(x) J^x ,  \\ J^0=\frac{i(uk \omega_0 b(x) +\omega u \partial_x)J^x}{u^2 k^2-\omega^2}= \frac{i \omega_0(uk -\omega \lambda \alpha)}{u^2 k^2-\omega^2} b(x) J^x \equiv i A^0 b(x) J^x. 
\end{align}
The introduced factors $A^y$ and $A^0$ can be interpreted as relative amplitudes for the $J^y$ and $J^0$ profiles with respect to the $J^x$ profile. Despite the presence of a singular denominator, the wave vector dependence is smooth. In terms of these amplitudes, the wave function for the Yanai MP mode can be presented as
\begin{equation}
\label{WFYanai}
\psi_\mathrm{Y}\propto\begin{pmatrix}
\frac{1-A^y b(x)}{\sqrt{2}} \\
i A^0 b(x)  \\
\frac{1+A^y b(x)}{\sqrt{2}}  \\
\end{pmatrix} \frac{1}{\left[\cosh\left(\frac{x}{d}\right)\right]^{\lambda}}.
\end{equation}
The transverse current profile $J^x$ is symmetric across the domain wall, but the $J^y$ and $J^0$ profiles have antisymmetric shapes. This behavior ensures that Kelvin and Yanai modes hosted by the domain wall are orthogonal to each other, $\langle\psi_{\mathrm{K}}|\psi_{\mathrm{Y}}\rangle=0$. This behavior also plays an important role in the hybridization between equatorial MP modes hosted by opposite facets in the cylinder geometry.

\begin{figure}[t]
	\begin{center}
		\includegraphics[trim=1.6cm 9cm 16cm 4.5cm, clip, width=0.45\columnwidth]{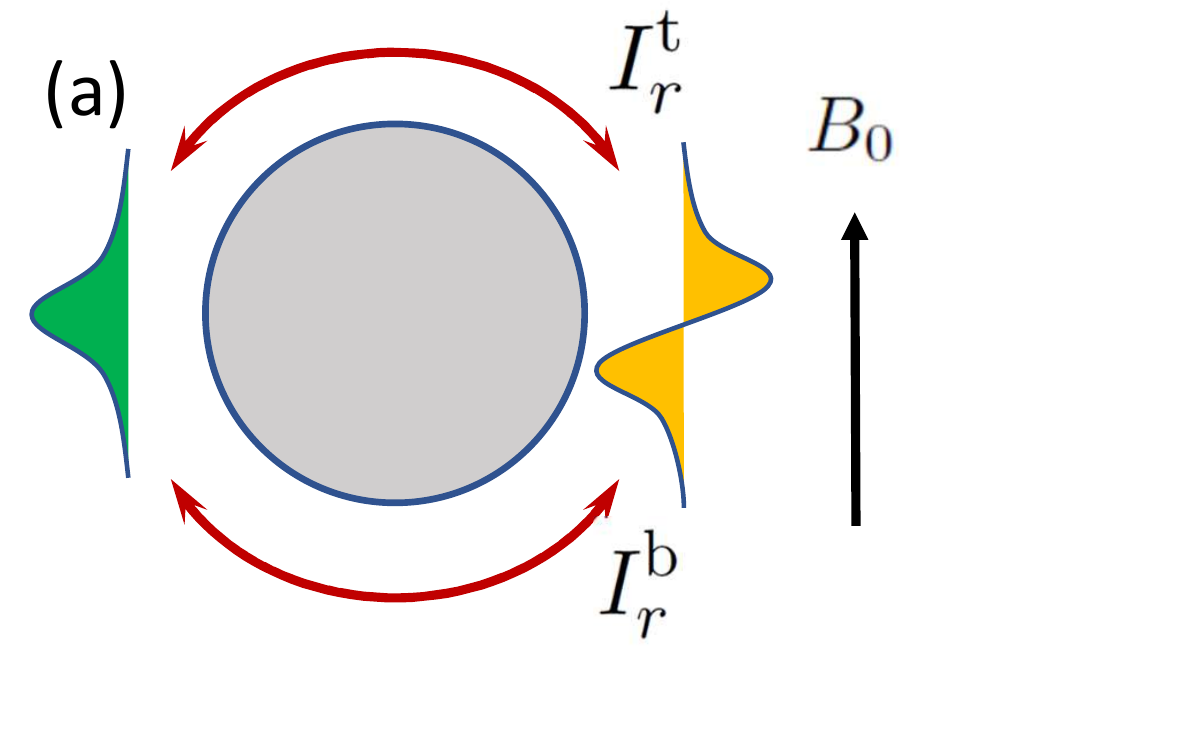}
		\includegraphics[trim=0cm 0cm 0cm 0cm, clip, width=0.45\columnwidth]{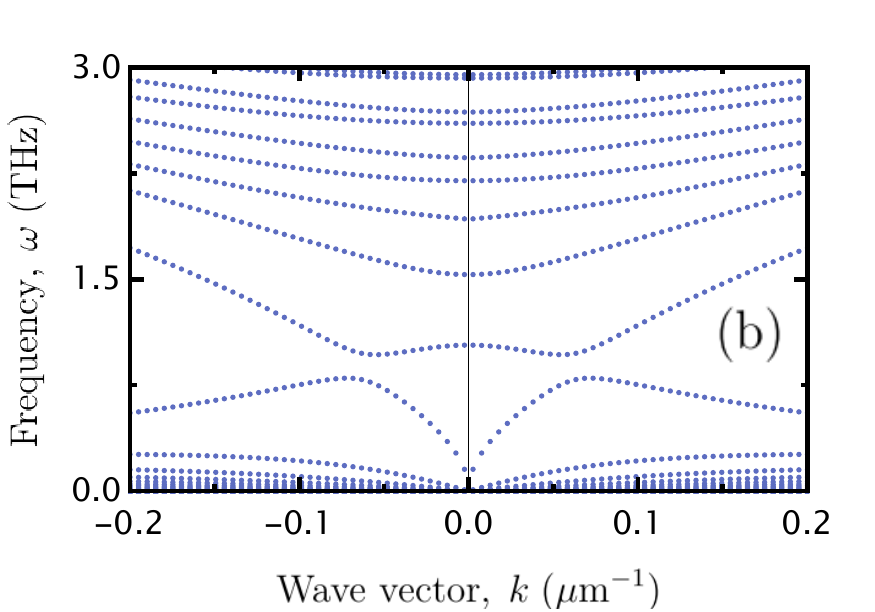}
		\caption{(a) The sketch illustrates the absence of the intermode hybridization due to the destructive interference between couplings across top and bottom hemicyliders.  This behavior arises from the fact that the transverse current $J^y$ and density $J^0$ components are symmetric for the Kelvin mode (green) and antisymmetric for the Yanai mode (yellow). (b) Spectrum for MP waves in the presence of a superimposed circularly symmetric magnetic field $B_\mathrm{C}=0.4 B_0$ with $R=10\;\mu\hbox{m}$. The exact cancellation of intermode hybridization due to destructive interference is lost. Kelvin and Yanai modes from different facets hybridize, and the resulting gap between hybrid modes is the most prominent in the vicinity of the avoided crossings.}
		\label{Fig5}
	\end{center}
	\vspace{-0.35in}
\end{figure}

\vspace{-0.1in}
\subsection{A2. Intermode hybridization and destructive interference effects}
\vspace{-0.05in}
For the case of a uniform magnetic field, the component that is perpendicular to the surface, $B^\mathrm{R}_{\vec{r}}=B_0 \cos(\theta)$, vanishes at opposite facets of the cylinder. As a result, facets can be seen as a pair of domain walls with opposite profiles (e.g., $b(x)$ and $b(x)'=-b(x)$), and the finite-size effects manifest as intermode hybridization. We address the hybridization effects with the help of the analytical model for MP waves introduced in Sec.~III of the main paper. 

This hybridization is especially important in the vicinity of the intersection ($\omega_{\mathrm{K}}'=\omega_{\mathrm{Y}}$ and $\omega_{\mathrm{K}}=\omega_{\mathrm{Y}}'$) between dispersion curves for modes hosted at different domain walls. The intersection point can be found analytically, $\omega_\star=\omega_0 \sqrt{\alpha/\alpha+2}$ and  $k_\star=\omega_\star/u$, and the corresponding wave functions for Kelvin and Yanai modes are given by

\begin{equation}
\psi_\mathrm{K}'\propto\begin{pmatrix}
- \frac{i}{\sqrt{2}} \\
-1  \\
\frac{i}{\sqrt{2}}  \\
\end{pmatrix} \frac{1}{\left[\cosh\left(\frac{x}{d}\right)\right]^{\frac{1}{\alpha}}},   \quad \quad \psi_\mathrm{Y}\propto\begin{pmatrix}
\frac{1-A^y_\star b(x)}{\sqrt{2}} \\
i A^0 b(x)  \\
\frac{1+A^y_\star b(x)}{\sqrt{2}}  \\
\end{pmatrix} \frac{1}{\left[\cosh\left(\frac{x}{d}\right)\right]^{\frac{1}{\alpha}}}, \quad \quad A^y_\star=-\frac{\alpha+1}{\sqrt{\alpha(\alpha+2)}}, \quad  A^0_\star=\frac{1}{\sqrt{\alpha(\alpha+2)}}.    
\end{equation}

The cylinder surface can be unfolded to a stripe with width $2\pi R$. The problem needs to be supplemented by the periodic boundary conditions and there is a pair of domain walls per period. As discussed below, the closed nature of the geometry is essential. However, it is instructive to begin the analysis with a pair of isolated domain walls. 

If we assume that the two domain walls are displaced by $2d_0$, the overlap between wave functions for the Kelvin and Yanai modes is given by 
\begin{equation}
\langle \psi'_\mathrm{K}| \psi_\mathrm{Y} \rangle =I_s\times I_r, \quad\quad I_s=i (A^y_\star+A^0_\star)=-i \sqrt{\frac{\alpha}{\alpha+2}}, \quad I_r=N_\mathrm{K} N_\mathrm{Y}\int_{-\infty}^\infty dx \; \frac{1}{\cosh^{\frac{1}{\alpha}}(\frac{x-d_0}{d})} \times \frac{\tanh(\frac{x+d_0}{d})}{\cosh^{\frac{1}{\alpha}}(\frac{x+d_0}{d})}. 
\end{equation}
Here, $N_{\mathrm{K}}$ and $N_{\mathrm{Y}}$ are normalization factors for the two modes, and their cumbersome explicit expressions are of importance here. The overlap between the Kelvin and Yanai modes is finite. As a result, the hybridization between modes in a system with two domain walls leads to a gap opening and an intersection between dispersion curves. 

Due to the closed nature of the cylindrical geometry, the hybridization is mediated across the top and bottom hemispheres. As a result, the overlap integral $I_r$ must be modified as $I_r\rightarrow I_r^\mathrm{t}+I_r^\mathrm{b}$, where $I_r^{\mathrm{t}}$ and $I_r^{\mathrm{b}}$ are given by     
\begin{equation}
I_r^\mathrm{t}=N_\mathrm{K} N_\mathrm{Y}\int_{-\pi R}^0 dx \; \frac{1}{\cosh^{\frac{1}{\alpha}}(\frac{x+\pi R}{d})}\times \frac{\tanh(\frac{x}{d})}{\cosh^{\frac{1}{\alpha}}(\frac{x}{d})}, \quad \quad \quad I_r^\mathrm{b}=N_\mathrm{K} N_\mathrm{Y}\int_{0}^{\pi R} dx \; \frac{1}{\cosh^{\frac{1}{\alpha}}(\frac{x-\pi R}{d})}\times \frac{\tanh(\frac{x}{d})}{\cosh^{\frac{1}{\alpha}}(\frac{x}{d})}.
\end{equation}
These terms have the same magnitudes but opposite signs, causing the total coupling $I_r$ to vanish. This behavior arises from the fact that the transverse current $J^y$ and density $J^0$ components are symmetric for the Kelvin mode and antisymmetric for the Yanai mode. In other words, the couplings across the top and bottom hemicylinders interfere destructively, thus forbidding intermode hybridization. This lack of intermode hybridization ensures the exceptional robustness of equatorial MP modes and their ability to overcome finite-size effects.

\vspace{-0.1in}
\subsection{A3. Breaking the destructive interference via an additional circularly symmetric magnetic field}
\vspace{-0.05in}
The cancellation of intermode hybridization relies on the symmetry between contributions across the top $I_r^\mathrm{t}$ and bottom $I_r^\mathrm{b}$ hemicylinders. If we superimpose uniform $B_0$ and circularly symmetric $B_\mathrm{C}$ magnetic fields, the corresponding radial component is given by $B_\vec{r}^\mathrm{R}=B_0 \cos (\theta)+B_\mathrm{C}$, and the symmetry between the hemicylinders is broken. The MP wave spectrum evaluated for $B_\mathrm{C}=0.4 B_0$ and $R=10\;\mu\hbox{m}$ (all other parameters are the same as in the main part of the paper) is presented in Fig.~5-b. As anticipated, the Kelvin and Yanai modes from different facets hybridize, and the resulting gap between the hybrid modes is most prominent in the vicinity of the avoided crossings. 

\vspace{-0.1in}
\section{APPENDIX B. MP waves in the spherical geometry}
\vspace{-0.05in}
\subsection{B1. Spherical scalar and vector harmonics}
For the spherical geometry, the system of integro-differential equations describing MP waves, Eqs.~(\ref{Continuity}) - (\ref{Potential}), can be reduced to algebraic equations via decomposition over the spherical vector harmonics. The decomposition is applied as follows: 
\begin{equation}
\begin{split}
\rho(\vec{r},t)=\sum_{n} \rho(n,t) Y_{n}(\vec{o}), \quad \quad \vec{j}(\vec{r},t)=\sum_n\left[j^Y(n,t) \vec{Y}_n(\vec{o})+ j^{\Psi}(n,t) \vec{\Psi}_n(\vec{o})+j^{\Phi}(n,t) \vec{\Phi}_n(\vec{o}) \right], \\
\phi(\vec{r},t)=\sum_{n} \phi(n,t) Y_{n}(\vec{o}), \quad \quad \vec{E}(\vec{r},t)=\sum_n\left[E^Y(n,t) \vec{Y}_n(\vec{o})+ E^{\Psi}(n,t) \vec{\Psi}_n(\vec{o})+E^{\Phi}(n,t) \vec{\Phi}_n(\vec{o}) \right].
\end{split}
\end{equation}
Here, $\vec{r}=\{R \cos \phi \sin\theta , R \sin \phi \sin\theta, R \cos\theta\}$ is the radius vector constrained to the surface of a sphere with radius $R$. For the sake of brevity, we have combined azimuthal $\theta$ and polar $\phi$ angles as $\vec{o}=\{\theta,\phi\}$. The index $n=\{l,m\}$ includes two angular discrete numbers, $l=0,1,2,...$ and $m=-l,...,l$. The function $Y_n(\vec{o})$ is the (scalar) spherical harmonic, which defines the vector spherical harmonics as follows:
\begin{equation}
\vec{Y}_n(\vec{o})=\vec{e}_\vec{r}^\mathrm{R} Y_{n}(\vec{o}), \quad \quad \vec{\Psi}_n(\vec{o})= \frac{R \vec{\nabla} Y_{n}(\vec{o})}{\sqrt{l(l+1)}}, \quad \quad \vec{\Phi}_n(\vec{o})= \frac{R [\vec{e}_\vec{r}^\mathrm{R} \times \vec{\nabla}Y_{n}(\vec{o})]}{\sqrt{l(l+1)}}.
\end{equation}
Here, $\vec{e}_\vec{r}^\mathrm{R}$ is the unit vector perpendicular to the surface of the sphere. For each value of $n$, the vector harmonics are orthogonal in the typical three-dimensional manner. In addition, the harmonics form a complete and orthonormal vector set, which is properly normalized as 
\begin{align}
\int d\vec{o}\; \vec{Y}_n^*(\vec{o}) \vec{Y}_{n'}(\vec{o})&=\delta_{n n'}, \quad \quad   \int d\vec{o}\; \vec{Y}_n^*(\vec{o}) \vec{\Psi}_{n'}(\vec{o})=0, \quad \quad \int d\vec{o}\; \vec{Y}_n(\vec{o})^* \vec{\Phi}_{n'}(\vec{o})=0,\\  
\int d\vec{o}\; \vec{\Psi}_n^*(\vec{o}) \vec{Y}_{n'}(\vec{o})&=0, \quad \quad   \int d\vec{o}\; \vec{\Psi}^*_n(\vec{o}) \vec{\Psi}_{n'}(\vec{o})=\delta_{n n'}, \quad \quad \int d\vec{o}\; \vec{\Phi}_n^*(\vec{o}) \vec{\Phi}_{n'}(\vec{o})=\delta_{n n'}.
\end{align}
Before addressing the effect of a uniform magnetic field on the MP wave spectrum, it is instructive to consider a spherically symmetric magnetic field profile. 
\vspace{-0.1in}
\subsection{B2. Spherically symmetric magnetic field}
\vspace{-0.05in}
A spherically symmetric magnetic field profile $B^\mathrm{R}(\vec{r})=B_0$ is just a useful model because it requires a magnetic monopole at the center of the sphere. Because the spherical symmetry is maintained, all spherical harmonics are decoupled as
\begin{align}
\label{SpericalLine1}
\partial_t \rho(\vec{r},t) +\vec{\nabla}\cdot \vec{j}(\vec{r},t)=0,\quad \quad \quad \quad \quad  &\longrightarrow \quad \quad \; \quad \quad   \partial_t\rho(n,t) = \frac{\sqrt{l(l+1)} J^\Psi(n,t)}{R}, \\ \label{SpericalLine2}
\partial_\tau j(\vec{r},t)=\frac{ne^2}{m} \vec{E}(\vec{r},t) + \frac{e B_0}{mc}[\vec{j}(\vec{r},t)\times \vec{e}_\mathrm{R}(\vec{r})], \quad \quad &\longrightarrow \quad \;  \partial_t J^{\Psi(\Phi)}(n,t)=\frac{ne^2}{m} E^{\Psi(\Phi)}(n,t) \pm \frac{e B_0}{mc} J^{\Phi(\Psi)}(n,t),\\ \label{SpericalLine3}
E(\vec{r},t)=-\nabla \phi(\vec{r},t), \quad \quad \quad \quad \quad  \quad \;\; &\longrightarrow \quad  E^\Psi(n,t)=-\frac{\sqrt{l(l+1)}\phi(n,t)}{R}, \quad \quad E^\Phi(n,t)=0, \\ \label{SpericalLine3}
 \; \; \phi(\vec{r},t) =\int d\vec{r}' V(\vec{r}-\vec{r}') \rho(\vec{r}',t) \quad \quad \quad \quad \quad   &\longrightarrow \quad \quad \quad  \quad \quad  \quad \; \quad   \phi(n,t)= \frac{4\pi R \rho(n,t)}{\kappa (2l+1) }. 
\end{align}
As for the case of planar geometry, this system of equations can be presented as a Hermitian-Schr\"{o}dinger-like eigenvalue problem, $\omega \psi(n,\omega)=\hat{H}(n) \psi(n,\omega)$ with $\psi(n,\omega)=\{j^+(n,\omega), j^0(n,\omega), j^-(n,\omega)\}$. Here, we have introduced $j^\pm(n,\omega)=(j^\Psi(n,\omega)\pm i j^\Phi(n,\omega))/\sqrt{2}$ and $j^0(n,\omega)=\omega(n) \rho(n,\omega)/q(n)$ with
\begin{equation}
\label{SpectrumSphere1}
q(n)=\frac{\sqrt{l(l+1)}}{R}, \quad \quad \quad \omega_\mathrm{p}(n)=\sqrt{\frac{4\pi n e^2 l(l+1)}{\kappa m (2l+1)}}.
\end{equation}
The effective Hamiltonian $H(n)$ and its spectrum are given by 
\begin{equation}
\label{SpectrumSphere2}
\hat{H}(n)=\begin{pmatrix}
\omega_0 & - \frac{i \omega_\mathrm{p}(n)}{\sqrt{2}} & 0 \\
 \frac{i\omega_\mathrm{p}(n)}{\sqrt{2}} & 0 &  \frac{i \omega_\mathrm{p}(n)}{\sqrt{2}} \\
0 &  -\frac{i \omega_\mathrm{p}(n)}{\sqrt{2}}& - \omega_0\\
\end{pmatrix}, \quad \quad \quad \Omega^0(n)=0, \quad \quad \quad \Omega^\pm(n)=\pm \sqrt{\omega_\mathrm{p}^2(n)+\omega_0^2}. 
\end{equation}
Here, $\omega_0=eB_0/mc$ is the Larmor frequency, and the frequency $\omega_\mathrm{p}(n)$ introduced in Eq.~(\ref{SpectrumSphere1}) determines the frequency of the plasma modes in the absence of a magnetic field. The expression for the spectrum, Eq.~(\ref{SpectrumSphere2}), resembles the expression derived for the planar geometry, Eq.~(\ref{MPBulkDispersion}), but with a discrete wave vector $\vec{q}\rightarrow q(n)$. Due to the spherical symmetry, the frequency does not depend on $m$, which results in the degeneracy of MP modes with the factor $2l+1$.   

\vspace{-0.1in}
\subsection{B3. Uniform magnetic field}
\vspace{-0.05in}
For the case of a uniform magnetic field penetrating the sphere, the radial component $B^\mathrm{R}_{\vec{r}} =B_0 \cos(\theta)$ shows a smooth dependence on latitude and vanishes along the equator. Although the spherical symmetry is broken, the axial symmetry remains, and $m$ is still a good discrete number. For a uniform magnetic field, Eq.~(\ref{SpericalLine2}) must be modified as follows: 
\begin{align}
\partial_t J^\Psi(n,t)=\frac{ne^2}{m} E^\Psi (n,t)  + \frac{eB_0}{mc} \sum_{n'}\left[M^{\Psi \Psi}_{nn'}J^\Phi(n',t) - M^{\Psi \Phi}_{nn'}J^\Psi(n',t)  \right], \\ 
\partial_t J^\Phi (n,t)=\frac{ne^2}{m} E^\Phi (n,t) + \frac{eB_0}{mc} \sum_{n'}\left[M^{\Phi \Psi}_{nn'}J^\Phi (n',t) - M^{\Phi \Phi} _{nn'}J^\Psi(n',t)  \right].
\end{align}
The calculation of the corresponding matrix elements is cumbersome but straightforward and results in   
\begin{align*}
M^{\Psi \Psi}_{nn'}=\int d\vec{o}\; \cos\theta\; \vec{\Psi}_n^* \cdot \vec{\Psi}_{n'}= \delta_{mm'}\left[\delta_{l,l'-1}\mu^\mathrm{S}_{n}+\delta_{l,l'+1}\mu^\mathrm{S}_{n'} \right], \quad \quad M^{\Psi \Phi}_{nn'}=\int d\vec{o}\; \cos\theta\; \vec{\Psi}_n^* \cdot \vec{\Phi}_{n'}\cdot=i \delta_{nn'} \mu_{n}^\mathrm{A} \\ 
M^{\Phi \Phi}_{nn'}=\int d\vec{o}\; \cos\theta\; \vec{\Phi}_n^*\cdot \vec{\Phi}_{n'}= \delta_{mm'}\left[\delta_{l,l'-1}\mu^\mathrm{S}_{n}+\delta_{l,l'+1}\mu^\mathrm{S}_{n'} \right], \quad \quad M^{\Phi \Psi}_{nn'}=\int d\vec{o}\; \cos\theta\; \vec{\Phi}_n^*\cdot \vec{\Psi}_{n'}=-i \delta_{nn'} \mu_{n}^\mathrm{A}.
\end{align*}
As expected, harmonics with different $m$ values are uncoupled. Here, we have introduced
\begin{equation}
\mu_n^\mathrm{S} = \frac{\sqrt{(l+1)^2-m^2}}{l+1} \sqrt{\frac{l(l+2)}{(2l+1)(2l+3)}}, \quad\quad\quad \mu_n^\mathrm{A}=-\frac{m}{l (l+1)}.     
\end{equation}
After applying the transformations described in the previous subsection, we obtain the following Hermitian-Schr\"{o}dinger-like eigenvalue problem for the MP wave spectrum: 
\begin{align}
\begin{split}
\omega \psi(l,m,\omega)=\begin{pmatrix}
\mu_{l m}^\mathrm{A} & - \frac{i \omega_\mathrm{p}(l,m)}{\sqrt{2}} & 0 \\
 \frac{i\omega_\mathrm{p}(l,m)}{\sqrt{2}} & 0 &  \frac{i \omega_\mathrm{p}(l,m)}{\sqrt{2}} \\
0 &  -\frac{i \omega_\mathrm{p}(l,m)}{\sqrt{2}}& \mu_{l m}^\mathrm{A}\\
\end{pmatrix} &\psi(l,m,\omega) + \\ \begin{pmatrix}
\mu_{l m}^\mathrm{S} & 0 & 0 \\
 0 & 0 &  0 \\
0 &  0& -\mu_{l m}^\mathrm{S}\\
\end{pmatrix} & \psi(l+1,m,\omega)  + \begin{pmatrix}
\mu_{l-1, m}^\mathrm{S} & 0 & 0 \\
 0 & 0 &  0 \\
0 &  0& -\mu_{l-1, m}^\mathrm{S}\\
\end{pmatrix} \psi(\omega,l-1,m).
\end{split}
\end{align}
The numerical solution of this eigenvalue problem is straightforward, and the corresponding results are presented in the main part of this paper. 

\begin{figure}[t]
	\begin{center}
		\includegraphics[trim=0cm 0cm 0cm 0cm, clip, width=0.55\columnwidth]{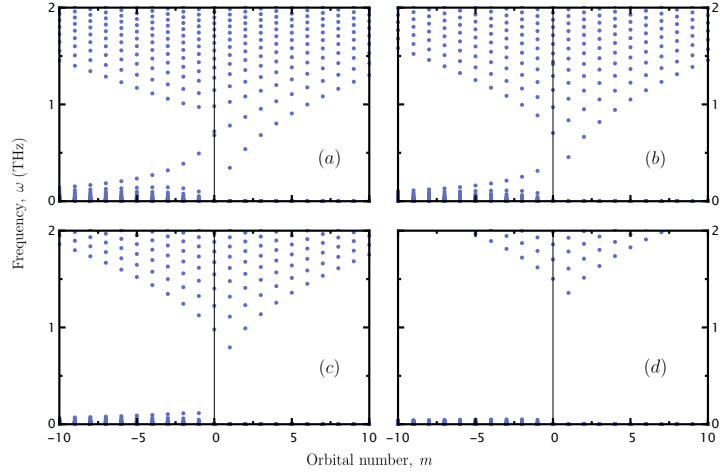}
		
		\caption{Spectrum of MP waves supported by a 2D electron gas confined at the surface of a sphere with radius $100\;\mu\hbox{m}$. The magnetic field includes superimposed spherically symmetric component $B_\mathrm{C}$ which is equal to $0.4 B_0$ (a), $0.8 B_0$ (b), $1.2 B_0$ (c) and $1.6 B_0$ (d). Here $B_0$ is magnitude of uniform magnetic field. For $B_\mathrm{C}>B_0$, Kelvin and Yanai modes disappear as expected, but the Rossby modes are still present.    
		}
		\label{Fig6}
	\end{center}
	\vspace{-0.3in}
\end{figure}

\vspace{-0.1in}
\subsection{B4. Rossby MP waves in the presence of superimposed spherically symmetric magnetic field} 
\vspace{-0.05in}

The mathematical origin of Kelvin and Yanai modes is intricately related with the topology of the bulk MP waves spectrum. As a result, their presence requires magnetic field to switch its sign across the equator. However, it is not the case for the Rossby waves. This can be demonstrated if we superimpose spherically symmetric and uniform magnetic fields. The corresponding radial component $B_\vec{r}^\mathrm{R}=B_0 \cos (\theta)+B_\mathrm{C}$ is latitude dependent, but does not involve the sign change if $B_\mathrm{C}>B_0$. The spectrum of MP waves is presented in Fig.~(\ref{Fig6}) for sphere radius $100\;\mu\hbox{m}$ (all other parameters are the same as in the main part of the paper) and superimposed magnetic field given by $0.4 B_0$ (a), $0.8 B_0$ (b), $1.2 B_0$ (c) and $1.6 B_0$ (d). As long as $B_\mathrm{C}<B_0$, the presence of Kelvin and Yanai modes are clearly visible, and their spectrum is weakly modified compared to $B_\mathrm{C}=0$. For $B_\mathrm{C}>B_0$, Kelvin and Yanai modes disappear as expected, but the topologically trivial Rossby modes are still present.

\end{appendix}
\end{widetext}

\end{document}